\documentclass{article}
\usepackage{arxiv}
\usepackage[utf8]{inputenc} 
\usepackage[T1]{fontenc}    
\usepackage{hyperref}       
\usepackage{amssymb}
\usepackage{amsmath}
\usepackage{url}
\usepackage{xcolor}
\usepackage{multirow}
\usepackage{booktabs}
\DeclareMathOperator*{\argmin}{arg\,min}
\usepackage{amsthm}
\usepackage{pdflscape}
\usepackage{textcomp}
\usepackage{caption}
\usepackage{subcaption}
\usepackage{comment}
\usepackage[inline]{enumitem}
\usepackage{url}
\usepackage{hyperref}
\usepackage[table]{xcolor}
\usepackage{adjustbox}
\usepackage{xr}

\definecolor{best}{RGB}{220,235,250}      
\definecolor{second}{RGB}{255,235,210}    
\definecolor{third}{RGB}{225,242,225}   

\title{GraphTFactor: A graph-based approach for proteins transcription factor prediction}

\author{
Domenico Amato \\
Department of Mathematics  and Computer Science\\
  University of Palermo\\
  Palermo, Italy 90123 \\
  \texttt{domenico.amato01@unipa.it} \\
   \And
Salvatore Calderaro \\
  Department of Mathematics  and Computer Science\\
  Department of Health Promotion, Mother and Child Care, Internal Medicine and  Medical Specialties \\
  University of Palermo\\
  Palermo, Italy 90123, 90127 \\
  \texttt{salvatore.calderaro01@unipa.it} \\
  \And
Giosuè  {Lo Bosco} \\
Department of Mathematics  and Computer Science\\
University of Palermo\\
Palermo, Italy 90123 \\
\texttt{giosue.lobosco@unipa.it} \\
 \And
Riccardo Rizzo \\
Institute for High Performance Computing and Networking\\
National Research Council of Italy\\
 Palermo, Italy 90146 \\
 \texttt{riccardo.rizzo@icar.cnr.it} \\
 \And
Filippo Vella\\
 Institute for High Performance Computing and Networking\\
 National Research Council of Italy\\
 Palermo, Italy 90146 \\
 \texttt{filippo.vella@icar.cnr.it} \\
}

\begin{document}
\maketitle
\begin{abstract}
Transcription factors (TFs) play an important role in regulating gene expression, making their identification a fundamental task in computational biology. In this work, we propose GraphTFactor, a novel end-to-end framework that models protein sequences as graphs and directly classifies transcription factors. By encoding sequences into graph structures, the proposed approach naturally handles variable-length inputs while preserving relevant relational information, without imposing length constraints. Experimental results show that GraphTFactor outperforms state-of-the-art methods on benchmark datasets, demonstrating the effectiveness of graph-based representations for TF prediction while providing interpretable insights into the sequence regions contributing to the model's predictions.
To support reproducibility and practical adoption, we release an open-source software package and an interactive web application that let users easily perform TF prediction on custom sequences.
\end{abstract}


\section{Introduction}
\label{sect:intro}
Transcription Factors (TFs) are DNA-binding proteins that play a fundamental role in regulating gene expression by controlling the initiation of transcription. 
Under specific physiological conditions, TFs selectively bind to certain DNA segments in the promoter region to control the expression levels of downstream genes \cite{karin1990too,latchman1997transcription}. The most common way to identify transcription factors is through biochemical experiments, such as specific DNA-binding assays \cite{helwa2010analysis}. However, conducting this kind of investigation is very labor-intensive, costly, and challenging, particularly on a large scale. To overcome these limitations, computational methods based on classical Machine Learning (ML) models have been developed \cite{eichner2013tfpredict,zheng2008combination}. Eichner et al. have introduced TFPredict and SABINE \cite{eichner2013tfpredict}, two sequence-based tools for predicting structural and functional properties of transcription factors. TFPredict relies on ML models and engineered features, including Position-Specific Scoring Matrix (PSSM) amino acid composition and recurring patterns, to distinguish TFs from non-TFs and assign functional families.
SABINE extends this analysis by including a model for predicting DNA-binding domains, structural features, and potential regulatory functions. Using these approaches, the authors have demonstrated that integrating evolutionary information and sequence-derived features significantly improves accuracy compared to homology-based methods. Zheng et al. \cite{zheng2008combination} have introduced a method for transcription factor identification and classification by combining Support Vector Machines (SVMs) with Error-Correcting Output Codes (ECOC). The approach relies on sequence-derived features—such as amino acid composition and physicochemical properties—without using alignment-based or homology-driven methods. Individual SVMs operate as binary classifiers, while ECOC encodes the multiclass problem into a set of binary tasks and decodes the outputs to achieve robust classification. This combination improves error tolerance and enhances both TF detection and family-level classification accuracy. This work represents an early and effective application of alignment-free machine learning strategies for TF prediction based solely on protein sequences.  Nevertheless, these models require a characteristic selection procedure that often involves computationally expensive homology measurements in biological sequences through alignment algorithms. In addition, TFs without homology to known TFs are very difficult to predict. 

In general, TF prediction represents a specific instance of sequence classification, and numerous deep learning (DL) methods have been introduced to address this problem, taking advantage of their inherent capability for automatic feature extraction \cite{amato2020deep1,Amato2021bookdeep,amato2020deep}. For the specific problem of TF prediction, a deep learning model named DeepTFactor was proposed in \cite{kim2021deeptfactor}. In this approach, proteins are encoded using one-hot representations, which are then processed by three parallel convolutional networks. The outputs of these branches are concatenated and fed into a classifier that determines whether the input sequence corresponds to a transcription factor or not. 
Although the method achieved excellent performance on both Eukaryotic and Prokaryotic protein datasets, its main characteristic lies in this fixed-length constraint. In this framework, all proteins are encoded as fixed-size matrices of $21 \times 1000$, where sequences shorter than 1000 amino acids are extended via zero-padding. Consequently, the proportion of zero entries in the representation is strongly dependent on the original protein length. This preprocessing step may therefore introduce a length-dependent signal that the classifier can exploit. Because proteins of different lengths generate systematically different input patterns, as well as due to the varying amount of padding required to reach a fixed input size. 

Recent advances in graph representation learning have shown that representing data using a graph structure can be helpful both in unsupervised and supervised classification by using Graph Neural Networks (GNNs) \cite{calderaro2023breast,amato2024bacteria,amato2024semantic}. In our previous work \cite{amato2024proteins}, we have proposed an approach for protein TF prediction based on the representation of a protein sequence leveraging the so-called De Bruijn graphs and the subsequent usage of a GNN for performing the final classification. The approach yields excellent performance comparable to that obtained using DeepTFactor \cite{kim2021deeptfactor} and superior to that obtained using one-hot encoding representation and classical predictive models such as SVMs. In particular, this approach allows for processing proteins with arbitrary length ($\geq k$) where $k$ is the dimension of the $k$-mers without requiring padding or arbitrary truncation of the sequence. In this paper, we propose an approach inspired by our previous method, representing each protein as a graph. Specifically, we extend the methodology introduced in \cite{amato2024proteins} by improving the graph construction pipeline, making it independent of the $k$-mer size, and leveraging Large Language Models (LLMs) to obtain biologically informed embeddings of protein subsequences. 
This representation allows the model to capture complex structural and functional relationships between subsequences, without limitations imposed by fixed or arbitrarily defined segment lengths, while incorporating the rich evolutionary and contextual information encoded in LLM-derived embeddings.
Experimental results show that the proposed approach improves the overall classification performance, with particularly notable gains in sensitivity, indicating a better ability to correctly identify transcription factors compared to previous methods.
The remainder of the paper is organized as follows. In Section~\ref{sect:material_methods}, we describe the annotation procedure and the creation of the datasets used to assess our framework. We then introduce the main components of GraphTFactor, including the procedure for encoding a protein as a graph and the GNN's architecture adopted for the classification task. In Section~\ref{sect:results}, we provide a comprehensive analysis of the validation protocol, the experiments performed, and the results obtained. Section~\ref{sect:impactlength} presents a study on the impact of transcription factor length on predictive performance. Specifically, we investigate whether the length of the input protein sequence affects the ability to accurately discriminate TF from non-TFs. This analysis aims to assess the robustness of the proposed graph-based representation across proteins of different lengths and to determine whether the observed predictive performance is influenced by sequence-length-dependent characteristics.
In Section~\ref{sect:ablation}, we investigate the contribution of the main components of the proposed framework through a series of ablation experiments. In Section~\ref{sect:discussion}, we critically discuss the obtained results, highlighting the strengths and limitations of our approach in comparison with state-of-the-art methods addressing the same problem. Finally, Section~\ref{sect:concl} summarizes the main conclusions of the study and outlines possible directions for future research, including potential improvements and applications of the proposed framework.

\section{Material and Methods}
\label{sect:material_methods}
This section describes the datasets used in the experiments, along with the adopted annotation procedure, and provides an overview of the GraphTFactor framework.
\subsection{Datasets generation}
\label{sect:data}
We constructed four datasets from the Swiss-Prot database downloaded in April 2024. Protein sequences were annotated as TFs or non-TFs following the Gene Ontology (GO)-based protocol adopted in \cite{kim2021deeptfactor}. Specifically, a protein was considered a TF if it had a GO annotation for transcription factor activity, or if it had both DNA-binding-related and transcription-regulation-related GO annotations. Proteins were further categorized according to their organism type as eukaryotic, prokaryotic, or viral. Sequences longer than 1,000 amino acids and those containing non-standard amino acids were excluded, consistent with the preprocessing criteria adopted in \cite{kim2021deeptfactor}. Following the TF-to-non-TF sampling strategy described in \cite{kim2021deeptfactor,amato2024proteins}, we constructed four datasets with an approximately 1:3 ratio between TF and non-TF sequences. The resulting \textit{All}, \textit{Eukaryotic}, \textit{Prokaryotic}, and \textit{Virus} datasets contain 19,986/59,958, 9,494/28,482, 9,954/29,862, and 538/1,614 TF/non-TF sequences, respectively. Further details on the annotation procedure, filtering criteria, taxonomic categorization, and dataset generation are reported in Section~\ref{sect:suppdata} of the  Appendix.
\subsection{The GraphTFactor Framework}
GraphTFactor is a novel deep learning-based framework for accurately predicting whether a protein contains a TF. It encodes a protein sequence, represented by its primary structure (the amino acid sequence), into a graph that models the similarity between informative subsequences. The subsequences are obtained using a Minimum Description Length (MDL) approach that segments the sequence into informative contiguous subsequences 
according to a criterion that minimizes the overall description length, thereby capturing the most informative patterns. 

\begin{figure}[ht]
	\centering
	\includegraphics[width=1\linewidth]{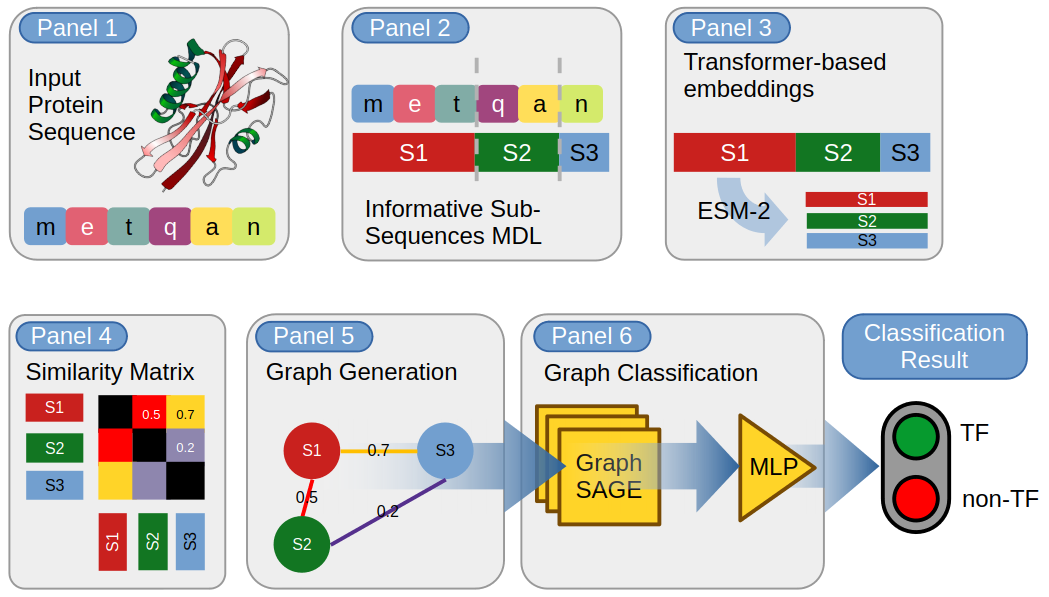}
	\caption{GraphTFactor pipeline: From a protein sequence, informative subsequences are embedded via ESM-2, encoded in a similarity-weighted graph, and analyzed by a GraphSAGE GNN and MLP classifier to predict transcription factor content.}
	\label{fig:pipeline}
\end{figure}

Then, the obtained segments are transformed into numerical representations using the Evolutionary Scale Modeling (ESM) model \cite{lin2023evolutionary}, which leverages large-scale protein language models to capture rich evolutionary and contextual information for each subsequence. The obtained embeddings are then used to construct a similarity matrix that guides the construction of graph topology. Once proteins have been encoded as graphs, they are fed into a GNN model based on GraphSAGE \cite{hamilton2017inductive}, which is trained in a supervised manner to predict whether the protein contains or does not contain a TF. During training, the model learns to aggregate information from neighboring nodes and generate informative graph embeddings, which are then used by a classifier to make the final predictions. A comprehensive illustration of the whole pipeline is depicted in Figure~\ref{fig:pipeline}. The following describes each step of the developed framework. 

\subsubsection{Encoding a protein as a graph}
Given a protein sequence $S$ of length $n$ on the amino acid alphabet $\Sigma$ with $\lvert \Sigma \rvert = 20$, we have segmented the sequence into $K$ contiguous segments using an approach based on the MDL:
\begin{equation}
	\label{eq:segment}
	S = \bigcup_{k=1}^K {S_k} \quad S_k=\left( s_{i_k},s_{i_k+1},\dots,s_{i_k+n_k}\right)
\end{equation}

where $S_k$ is a segment of $S$ with $\lvert S_k \rvert=n_k$. 
The segmentation is obtained by minimizing the following cost:
\begin{equation}
	\label{eq:cost}
	\mathcal{L}(S) = \sum_{k=1}^{K}{\left[ \mathcal{C}({S_k}) + \lambda \right]}
\end{equation}
where $\mathcal{C}(S_k)$ is the encoding cost of the segment $S_k$ and $\lambda=\log{n}$ is the cost for introducing a new segment. For the computation of the cost on an arbitrary segment $S_k$, we first calculate the frequency $c_a$ and probability $p_a$ of each amino acid $a \in \Sigma£$ within the segment in the following way:
\begin{equation}
	\label{eq:prob}
	c_a=\#\left\{ s_{l}=a \; : \; i_k \le l \le i_k+n_k\right\} \quad  \quad p_a = \frac{c_a}{n_k}.
\end{equation}
Using these probabilities, the encoding cost of the segment is computed according to the empirical Shannon entropy scaled by the segment length:
\begin{equation}
	\label{eq:entropy}
	\mathcal{C}(S_k) = - \sum_{a \in \Sigma_k}{c_a \log{p_a}}
\end{equation}
where $\Sigma_k$ is the amino acid alphabet of the considered segment. $\mathcal{C}(S_k)$ quantifies the amount of information required to encode the segment under the observed amino acid distribution. Segments with more uniform distributions or rare amino acids will have higher encoding costs, while segments with repetitive or predictable patterns will have lower costs. This formulation ensures that the MDL-based segmentation favors informative and non-redundant subsequences, which are subsequently used as nodes in the protein graph representation. To obtain the optimal segmentation, it is possible to use several strategies. 
In this work, we have employed a strategy based on the dynamic programming paradigm \cite{bellman1957dynamic}. A more detailed description of the employed strategy can be found in Section~\ref{sect:suppdp} of the Appendix. 
Once the sequence was segmented, the obtained protein subsequences were numerically encoded using a pretrained protein language model, specifically the ESM model \cite{rives2021biological}, which is a transformer-based model trained on large-scale protein sequence databases and learns rich contextual representations of amino acids by capturing both local and long-range dependencies within sequences. The Transformer mechanism relies on multi-head self-attention, allowing the model to dynamically weight interactions between all residues in the sequence, thereby capturing patterns related to protein structure, functional motifs, and evolutionary constraints. As a result, the learned embeddings provide rich representations that have been shown to correlate with structural and biochemical properties of proteins, making them particularly suitable for downstream predictive tasks.
These embeddings serve as a numerical representation of the sequence, providing a meaningful feature space for downstream tasks. In our study, we use the ESM-2 model \cite{lin2023evolutionary}, which extends the original ESM architecture with increased depth and parameter capacity, yielding more expressive embeddings that better capture subtle sequence patterns and relationships. 
To derive the embeddings of the subsequences, we have used the following approach. Firstly, we have computed the token embeddings of the whole sequence $S$, removing the \textit{CLS} and \textit{EOS} tokens, obtaining a matrix $X \in \mathbb{R}^{n \times d}$ where $n$ is the number of amino acids, and $d$ is the embedding size. 
Each row $x_i \in \mathbb{R}^d$ in $X$ represents the embedding of the i-th amino acid. Given a subsequence $S_k$ defined by the interval of indices $[b_k,e_k)$ with $1 \leq b_k < e_k \leq n$, we compute a fixed-dimensional representation by aggregating the embeddings of the amino acids belonging to that region. This aggregation is necessary because the subsequent graph construction operates at the subsequence level rather than at the amino acid level. Consequently, each variable-length subsequence must be summarized by a single embedding while preserving as much contextual information as possible.
Instead of assigning equal importance to all amino acids within a subsequence, we employ a weighted aggregation scheme that gives greater importance to the ones located closer to the segmentation boundary. This choice is motivated by the left-to-right nature of the segmentation procedure: the boundary marks the transition between the current subsequence and the subsequent region, and amino acids immediately preceding it are the last sequence elements considered before this transition. Consequently, these amino acids are expected to provide more informative contextual cues about the local sequence characteristics at the boundary and, potentially, about the properties of the following region. By assigning higher weights to amino acids closer to the boundary, the aggregation process can therefore preserve this positional information while reducing the contribution of amino acids located farther from the transition.
The weights are defined as
\begin{equation}
	\label{eq:weights}
	w_j = \frac{1}{2} + \frac{j-1}{n_k -1}\times\frac{1}{2}, \qquad j=1,\dots,n_k,
\end{equation}
and are normalized to ensure that they sum to one:
\begin{equation}
	\label{eq:norm_weights}
	\tilde{w}_j=\frac{w_j}{\sum_{l=1}^{n_k}w_l}.
\end{equation}
This normalization is important because it preserves the scale of the resulting embedding independently of the subsequence length. Without normalization, longer subsequences would naturally produce embeddings with larger magnitudes simply because they contain more weighted terms.
The embedding of subsequence $S_k$ is then computed as

\begin{equation}
	\label{eq:subseqemb}
	r_k=\sum_{j=1}^{n_k}\tilde{w}_j x_{b_k+j-1},
\end{equation}
where $r_k\in\mathbb{R}^d$ is the embedding representation of the subsequence. 
Although the subsequence embeddings capture the semantic content of individual regions, considering each embedding independently does not explicitly represent how the protein sequence changes from one region to the next. Since the subsequences are ordered according to their position along the protein sequence, we additionally encode the local variation between consecutive regions using a first-order difference representation.

Let $R\in\mathbb{R}^{K\times d}$ denote the matrix containing the embeddings of the $K$ subsequences, with $r_k\in\mathbb{R}^d$ representing the embedding of the $k$-th region. We define the difference-based representation as
\begin{equation}
	\label{eq:localdiff}
	\hat{x}_k =
	\begin{cases}
		r_1, & k=1,\\
		r_k-r_{k-1}, & k=2,\dots,K.
	\end{cases}
\end{equation}

For the first subsequence, no preceding region is available; therefore, its original embedding $r_1$ is retained. For each subsequent subsequence, the representation is instead given by the difference between its embedding and that of the immediately preceding region. Thus, while $r_k$ describes the characteristics of the $k$-th region, $r_k-r_{k-1}$ captures the direction and magnitude of the change occurring between two consecutive regions.

\begin{figure}
	\centering
	\includegraphics[width=0.8\linewidth]{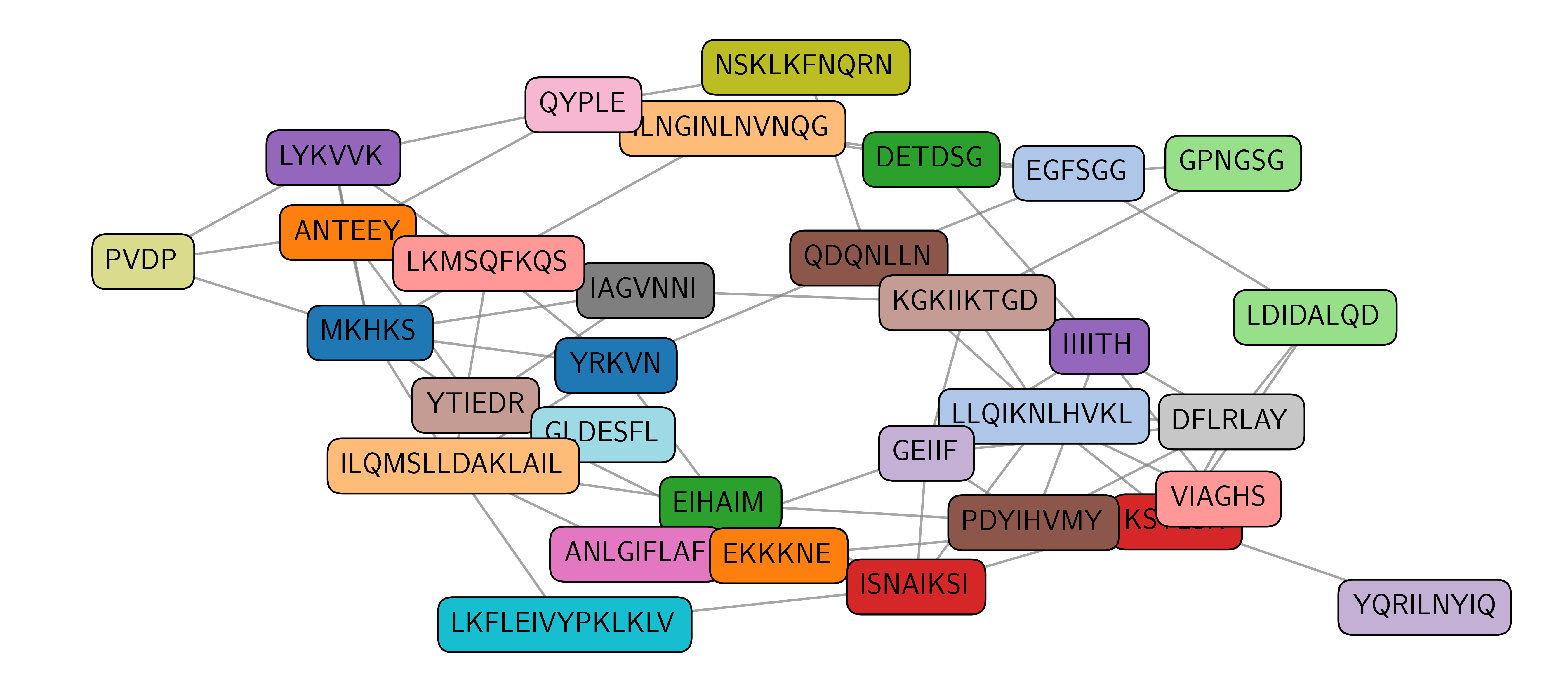}
	\caption{Example of the graph representation adopted in GraphTFactor. Each node corresponds to a subsequence in the protein sequence, while edges represent relationships between subsequences that the model uses to learn structural and contextual dependencies.}
	\label{fig:graph}
\end{figure}

The resulting matrix $\hat{X}\in\mathbb{R}^{K\times d}$ consequently combines absolute information for the first region with local transition information for all subsequent regions. This representation allows the model to distinguish not only which sequence features characterize each individual region, but also how these features evolve along the protein sequence. Such local changes are particularly relevant for graph construction, as they provide an additional representation of the sequential relationships between neighboring regions and allow the resulting graph to encode both regional characteristics and their local evolution along the protein sequence.
The illustration depicting ESM-2 embedding computation for the subsequences is reported in Panel 3 of Figure~\ref{fig:pipeline}.
Finally, to build the graph, we have computed a similarity matrix $D = \hat{X} \hat{X}^\top \in \mathbb{R}^{K \times K}$ where each entry $D_{ij}$ represents the inner product similarity between subsequences' embeddings. 
Note that, before computing pairwise similarities, each row of $\hat{X}$ is L2-normalized to unit norm:
\begin{equation}
	\label{eq}
	\hat{x}_k =
	\frac{\hat{x}_k}{\lVert \hat{x}_k \rVert_2}.
\end{equation}
This normalization ensures that all subsequence representations have the same magnitude, preventing differences in vector norm from dominating the similarity computation. Consequently, the subsequent inner product primarily reflects the directional similarity between subsequence representations, making the resulting similarity values comparable across different regions and proteins.
We then construct the edges of the graph by connecting pairs of nodes $(i,j)$ whose similarity $D_{ij}$ exceeds an adaptive threshold $\tau$ computed as the 70th percentile of the non-zero similarity values. The corresponding $D_{ij}$ values are used as edge weights, and self-loops are removed. The choice of the 70th percentile is motivated by the need to retain only the most significant similarities while avoiding overly dense graphs.
This results in a weighted graph representation $\mathcal{G} = (\mathcal{V}, \mathcal{E},\hat{X},W)$ of the sequence where:
\begin{itemize}
	\item $\mathcal{V}$ is the set of nodes (the subsequences);
	\item $\mathcal{E}$ is the set of edges (the interactions between subsequences);
	\item $\hat{X} \in \mathbb{R}^{K \times d}$ is the initial node features matrix (the embedding of each subsequence); the dimensionality $d$ depends on the ESM-2 model employed, as reported in Table~\ref{table:esm_conf};
	\item $W \in \mathbb{R}^{\lvert \mathcal{E} \rvert}$ is the vector containing the edge weights. 
\end{itemize}
The procedure illustrating the overall mechanism of graph construction is reported in Panels 4 and 5 of Figure~\ref{fig:pipeline}, while an example of a protein encoded as a graph through this procedure is shown in Figure~\ref{fig:graph}. In this representation, each node corresponds to a subsequence of the original protein sequence, while edges are established only between nodes showing strong similarity relationships.

\subsubsection{Classification using Graph Neural Networks}
Once the proteins were encoded as graphs, they were used to train a GNN to predict whether a protein contains a TF. GNNs are a class of neural networks specifically designed to process graph-structured data by learning representations that capture both node features and graph topology. They can be applied to various tasks, including node classification, link prediction, and graph classification. In this study, TF prediction was formulated as a graph-level classification problem, where each protein graph is assigned a binary label indicating the presence or absence of a TF. In graph classification, a readout function is required to aggregate node-level embeddings into a single graph-level representation. This readout operation (e.g., global mean pooling, sum pooling, or max pooling) produces a fixed-size vector that is subsequently passed to a classifier to perform the final prediction. Given the graph $\mathcal{G} = (\mathcal{V}, \mathcal{E},\hat{X}, W)$, the aim of the GNN is to learn a new node representation $H \in \mathbb{R}^{\lvert V \rvert \times F}$ where $F$ is the dimensionality of the learned features for each node \cite{xu2018how}. This new representation is obtained by leveraging graph topology, edge weights, and initial node features. 

A GNN consists of a series of stacked layers $l=1,\dots,L$, and works iteratively, updating the node representation by combining it with the ones of its neighborhood.
In each layer $l$, two processing steps are performed: an \textit{aggregation} step that combines (often sums) the features of the neighborhoods, and a \textit{combine} step that updates the current node features using the information from the previous step.

The proposed architecture is a GNN for graph-level prediction built upon the GraphSAGE message-passing framework \cite{hamilton2017inductive}. The model first learns node representations through three stacked GraphSAGE convolutional layers that progressively transform the input node features into higher-level embeddings with dimensions 256, 128, and 64, respectively, where 64 represents the output node features dimensionality. Each convolutional layer is followed by batch normalization \cite{ioffe2015batch}, a ReLU activation function, and dropout regularization in order to stabilize training and mitigate overfitting. Note that, since the constructed graphs are weighted, the edge weights are incorporated in the message-passing step by multiplying the feature vector of each neighboring node by the corresponding edge weight before the aggregation operation. In this way, the aggregated representation becomes a weighted mean of the neighboring node features, allowing stronger edges to contribute more to the updated node embeddings.
After the message-passing stage, we aggregate node embeddings into a graph-level representation using two complementary readout operations: global max pooling and global mean pooling. Each pooling operation produces a 64-dimensional vector, which we then concatenate to obtain a 128-dimensional graph-level representation. This combination captures both the most salient node-level features and the average structural characteristics of the graph (see Panel 6 of Figure~\ref{fig:pipeline}).
This representation is then processed by a multilayer perceptron comprising three fully connected layers with dimensions 128–64–32–1, with ReLU activations and dropout applied between the hidden layers. The final layer produces a single scalar output corresponding to the prediction associated with each input graph (see Panel 7 of Figure~\ref{fig:pipeline}). The network was trained for 100 epochs using the Adam optimizer \cite{KingmaB14} with a learning rate of $0.001$  and a minibatch size of 64. 

\section{Results}
\label{sect:results}
\begin{figure}
	\centering
	\begin{subfigure}[b]{0.49\textwidth}
		\centering
		\includegraphics[width=\textwidth]%
		{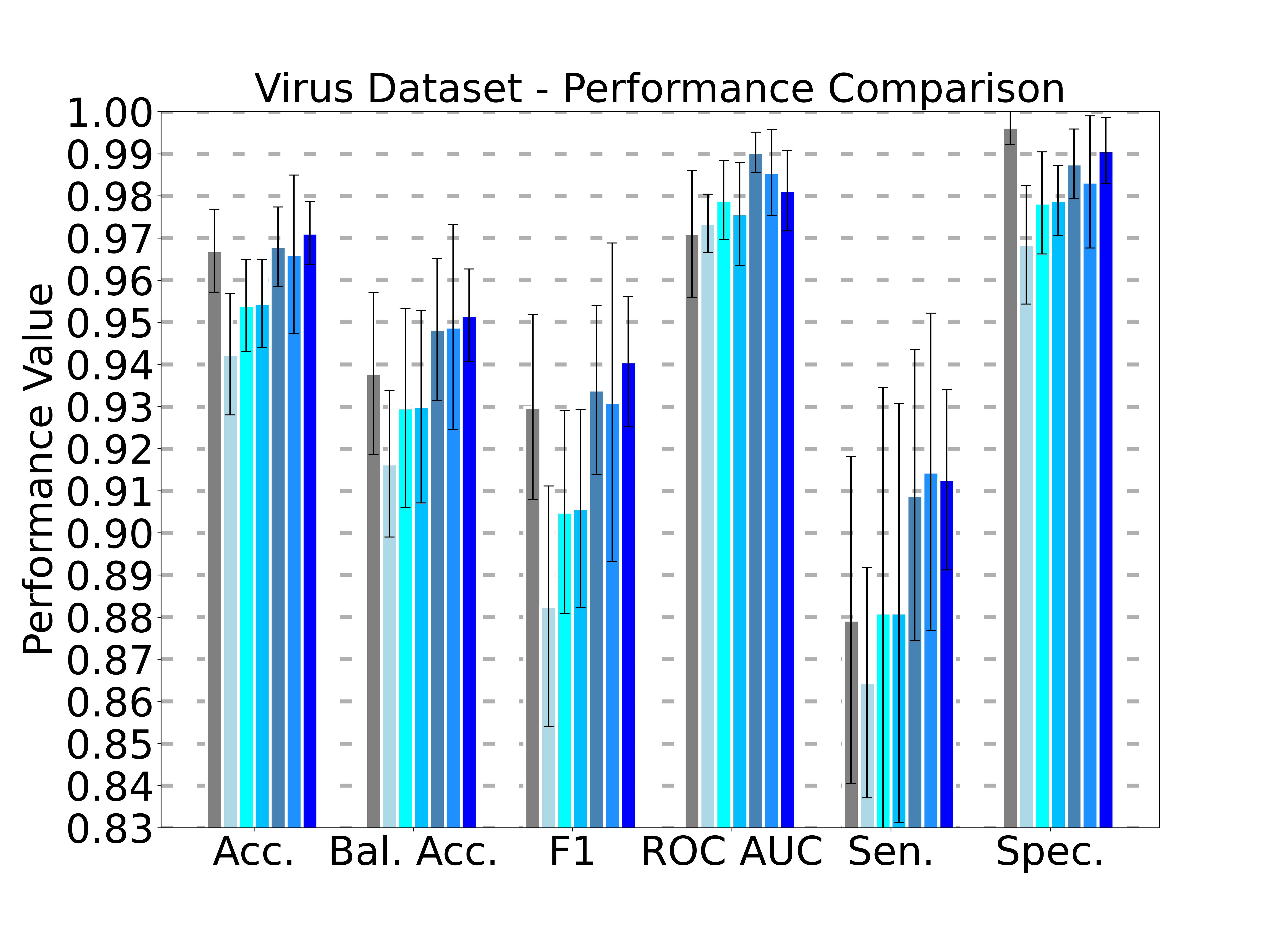}
		\label{fig:virus_hist}
	\end{subfigure}
	\hfill
	\begin{subfigure}[b]{0.49\textwidth}
		\centering
		\includegraphics[width=\textwidth]{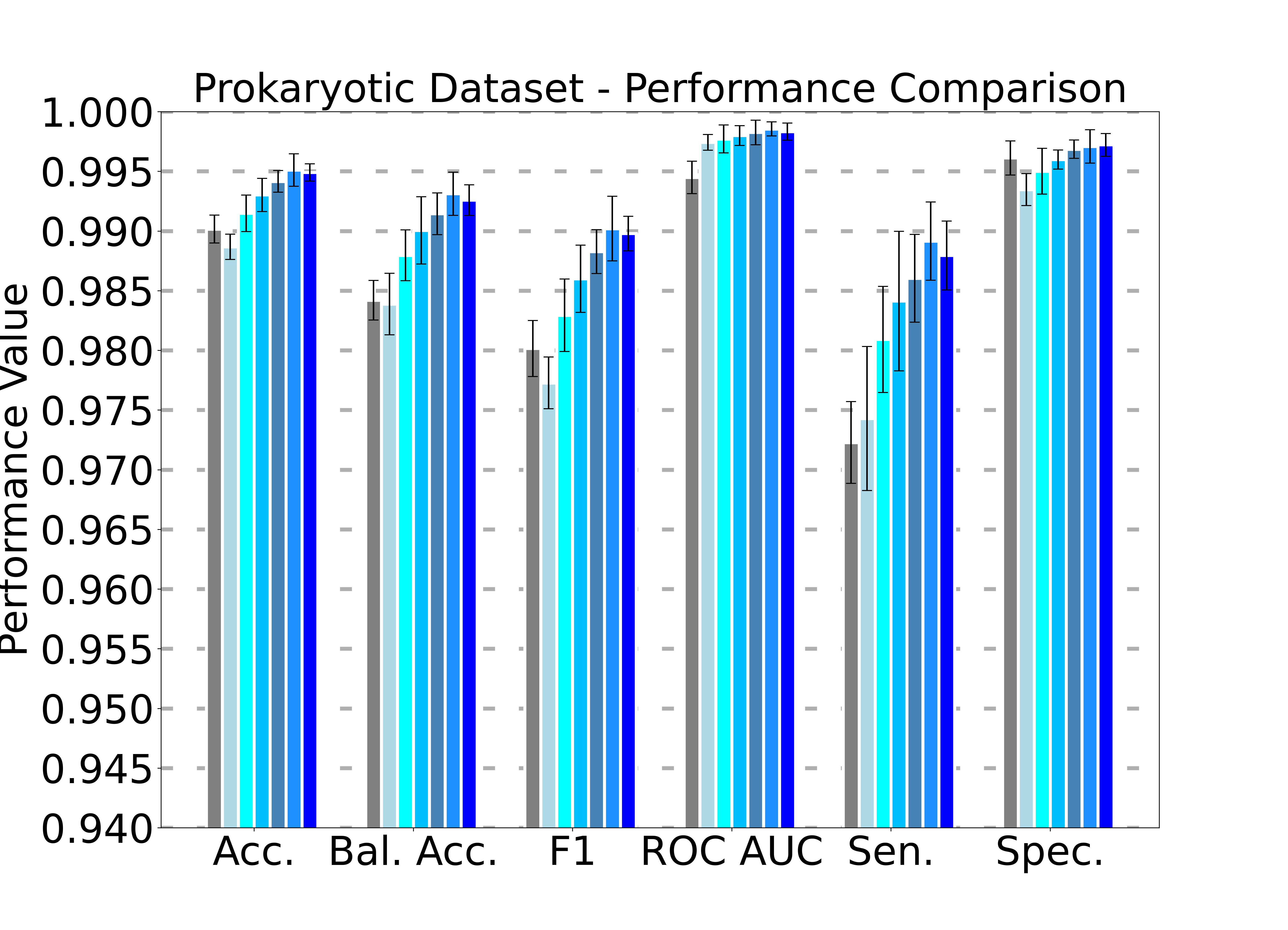}
		\label{fig:Proka_hist}
	\end{subfigure}
	\begin{subfigure}[b]{0.49\textwidth}
		\centering
		\includegraphics[width=\textwidth]{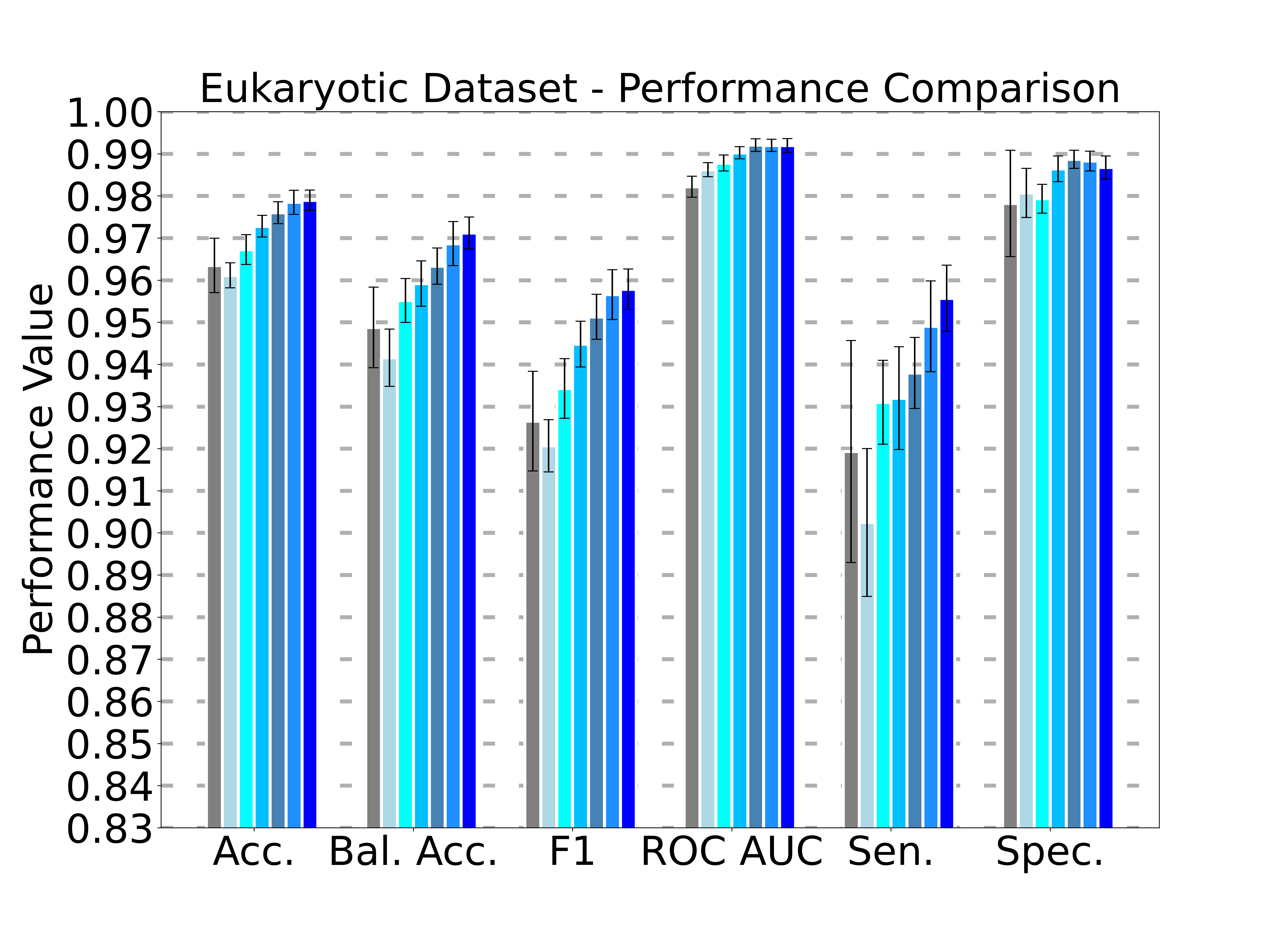}
		\label{fig:Euka_hist}
	\end{subfigure}
	\hfill
	\begin{subfigure}[b]{0.49\textwidth}
		\centering
		\includegraphics[width=\textwidth]{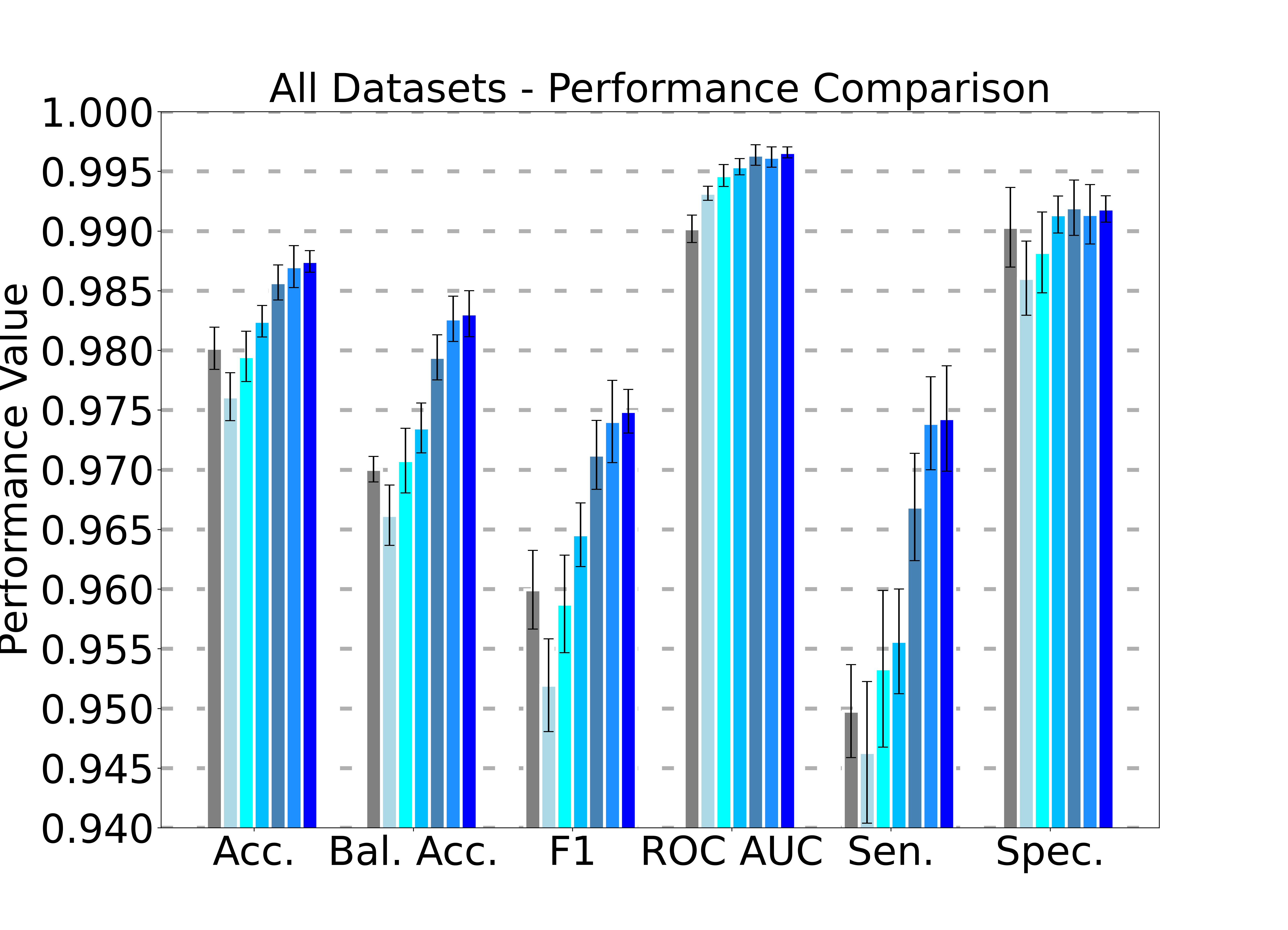}
		\label{fig:All_hist}
	\end{subfigure}
	
	\begin{subfigure}[b]{0.99\textwidth}
		\centering
		\includegraphics[width=\textwidth]{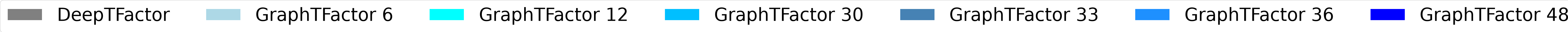}
		\label{fig:legend_hist}
	\end{subfigure}
	
	\caption{Histogram comparison across the \emph{Virus}, \emph{Prokaryotic}, \emph{Eukaryotic}, and \emph{All} datasets.}
	\label{fig:three graphs}
\end{figure}

In this section, we present and analyze the outcomes of the experimental evaluation of the GraphTFactor framework. Its performance was systematically compared against the state-of-the-art deep learning model DeepTFactor \cite{kim2021deeptfactor}, employing a 10-fold stratified cross-validation protocol. The adoption of a stratified cross-validation strategy is motivated by its capacity to preserve the empirical class or label proportions of the original dataset, thereby providing a more realistic approximation of the true data-generating distribution.
For each fold, we have computed key performance metrics: accuracy (Acc), balanced accuracy (BA), sensitivity (Se), specificity (Spe), F1-Score (F1), and area under the ROC curve (AUC), reporting both the mean and standard deviation to assess the reliability and consistency of the results. All experiments were performed on a workstation equipped with 12th Gen Intel\textsuperscript{\textregistered} Core™ i7-12700F × 20 and an NVIDIA RTX 6000 Ada Generation. The total amount of system memory is 128 GB of DDR5. The GPU is also equipped with 48 GB of GDDR6 VRAM and uses the CUDA parallel computing platform. The operating system is Ubuntu 24.04.4 LTS. The experimental evaluation involved testing the proposed GraphTFactor framework on the four datasets described in subsection~\ref{sect:data}, and comparing its performance with DeepTFactor using the same hyperparameters reported by the authors in \cite{kim2021deeptfactor}. For GraphTFactor, we evaluated several variants of the framework obtained by using different ESM-2 models to compute the subsequence embeddings. The considered models and their main characteristics are reported in Table~\ref{table:esm_conf}. Specifically, the first column indicates the model variant, identified by the number of layers in the network; the second column reports the total number of parameters of the model; and the third column shows the dimension of the embedding produced by the model, denoted as $d$. As shown in the table, the considered configurations differ in model capacity and representational power: deeper models generally produce higher-dimensional embeddings and have more parameters, which may enable the extraction of richer sequence features at the cost of increased computational requirements. Note that, for the ESM-2 model, we do not perform any form of fine-tuning or model re-training, but we use it directly to derive the subsequences' numerical embeddings. 
The experimental results and the comparison with DeepTFactor are summarized in Figure~\ref{fig:three graphs}, which reports, for each experiment, the mean and standard deviation of the evaluation metrics computed across the 10 folds of the cross-validation. The complete numerical results are reported in Table~\ref{table:results} of the Appendix.
Across all datasets, GraphTFactor consistently demonstrates competitive or superior performance, and architectures leveraging initial node embeddings with higher dimensions lead to more robust and reliable predictions. For the \textit{All} dataset, GraphTFactor progressively improves over DeepTFactor as the number of layers of the ESM-2 model increases. While DeepTFactor provides strong baseline performance, GraphTFactor benefits from its graph-based representation, which captures complex relationships within sequences, thereby enhancing both sensitivity and specificity. This allows the model to make more balanced predictions and better differentiate between TF and no-TF proteins. For the \textit{Eukaryotic} dataset, which contains more complex and variable sequences, GraphTFactor clearly outperforms DeepTFactor. The variants of GraphTFactor using deeper ESM-2 architectures are particularly effective in identifying subtle sequence patterns that may be missed by the baseline method. For the \textit{Prokaryotic} dataset, both GraphTFactor and DeepTFactor achieve high overall performance, reflecting the relative simplicity and consistency of these data. Nevertheless, GraphTFactor exhibits incremental advantages, particularly in terms of prediction stability and reliability. This suggests that the graph-based approach can extract hierarchical and contextual features even in datasets where the sequences are more uniform, offering more consistent results across metrics. The \textit{Virus} dataset presents the greatest challenge due to high sequence variability and heterogeneity. Here, shallower GraphTFactor models initially underperform compared to DeepTFactor, but deeper architectures quickly surpass the baseline. The increased depth enables GraphTFactor, resulting in improved sensitivity and more balanced predictions. This demonstrates the model’s ability to adapt to highly diverse and challenging datasets, where traditional approaches may struggle. Taken together, these results show that GraphTFactor not only matches but often exceeds the performance of DeepTFactor across diverse biological contexts. By leveraging graph-based representations and deeper architectures, it improves prediction balance and provides more reliable and robust classification. These findings highlight the potential of GraphTFactor as a flexible and powerful tool for transcription factor prediction across a wide range of sequence types.

\begin{table}[ht]
	\centering
	\caption{Overview of ESM-2 model variants, showing the number of layers, total parameters, and embedding dimension ($d$) for each model size.}
	\label{table:esm_conf}
	\begin{tabular}{@{}lll@{}}
		\toprule
		\textbf{ESM-2 \#layers} & \textbf{Number of parameters} & \textbf{$\mathbf{d}$ (Embedding Size)} \\ \midrule
		6                    &       8M                        &     320                       \\
		12                   &       35M                       &     480                        \\
		30                   &       150M                     &      640                      \\
		33                   &       650M                     &      1280                     \\
		36                   &       3B                    &         2560                     \\
		48                   &       15B                  &          5120                     \\ \bottomrule
	\end{tabular}
\end{table}

To facilitate the practical use and accessibility of GraphTFactor, we developed and publicly released a dedicated Python package and a web application, providing complementary interfaces for researchers with different levels of programming expertise. Regarding the Python package \cite{amato2026graphtfactorpack}, we developed a dedicated utility that accepts custom protein sequences provided either as raw text or in FASTA format and enables (TF) prediction using all GraphTFactor variants presented in this manuscript. When available, users can also specify the organism of the input protein sequence, allowing the corresponding organism-specific model to be selected for prediction. The utility returns a data frame in which each row corresponds to an input protein and reports the predicted class together with the associated prediction confidence, expressed as the probability returned by the sigmoid output unit of the network. The resulting data frame can optionally be saved to disk by specifying an output path, allowing users to preserve the results for subsequent analyses and avoid repeating computationally demanding predictions.
The web application \cite{amato2026graphtfactorweb} provides a user-friendly interface through which users can paste the protein sequence to be analyzed, specify the organism type, when known, and select the desired GraphTFactor variant. Due to the computational constraints of the platform hosting the application, only the GraphTFactor variants based on the ESM-2 models with 6, 12, and 30 layers are currently available through the web interface.
Once the required information has been provided, clicking the Predict button initiates the complete analysis pipeline. The application then provides a graphical representation of the amino acid composition of the input sequence, followed by a visualization of the graph constructed from the sequence. Finally, the application displays the predicted class together with the corresponding confidence value, providing users with an intuitive overview of both the input sequence characteristics and the resulting TF prediction.

\section{Impact of sequence length on transcription factor prediction: a direct comparison between DeepTFactor and GraphTFactor}
\label{sect:impactlength}

\begin{figure}[ht]
	\centering
	\includegraphics[width=1\linewidth]{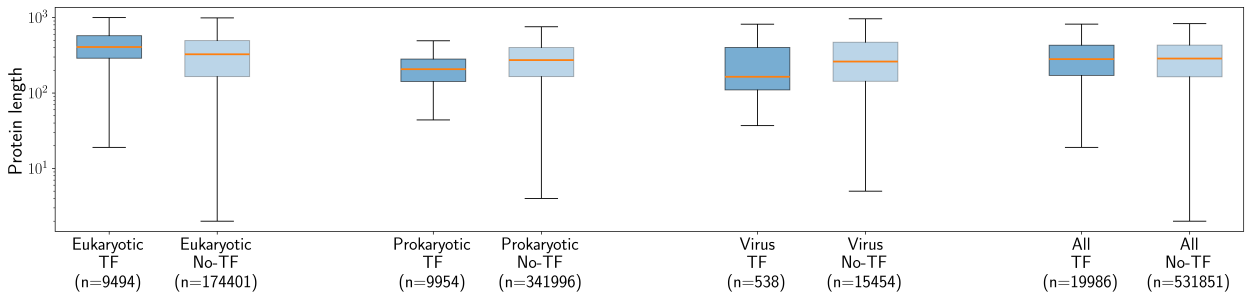}
	\caption{Distribution of protein sequence lengths for TFs and no-TFs across different organism groups.}
	\label{fig:boxplot}
\end{figure}

A potentially important aspect to investigate is whether protein sequence length itself contributes to TF prediction. This issue is particularly relevant for DeepTFactor, whose input representation is based on one-hot encoding.  As a result, the model may partially rely on sequence length or padding-related artifacts rather than exclusively learning biologically meaningful features associated with DNA-binding activity and transcriptional regulation. Figure~\ref{fig:boxplot} presents boxplots of protein sequence lengths for each organism category considered in this study. Although TFs tend to be slightly longer than non-TFs in eukaryotes, the distributions largely overlap.
To assess this effect, we systematically analyzed the relationship between protein length and prediction outcomes. The main objective is to determine whether the predictive performance of DeepTFactor is influenced by biases introduced by the fixed-length encoding scheme. 
To isolate the effect of sequence length from class imbalance, we constructed length-balanced datasets in which TFs and non-TFs share comparable length distributions. After removing sequences longer than 1000 amino acids and sequences containing non-standard residues, proteins were grouped into 50 bins. For each bin, an equal number of TF and non-TF proteins were randomly sampled, using the smaller class size as the sampling limit. This procedure yields datasets with matched length distributions between classes, enabling a more direct assessment of the contribution of sequence length to predictive performance.
A similar pattern is observed in prokaryotes and viruses, where differences in median length are modest relative to the substantial within-class variability. When all organisms are considered together, TFs show a slightly higher median length; however, the interquartile ranges remain highly overlapping, indicating that sequence length alone provides limited discriminatory power at the global level.

We next evaluated DeepTFactor and GraphTFactor on the length-balanced datasets to quantify the extent to which predictive performance depends on sequence length. For this kind of experiment, we used the same validation protocol described in Section~\ref{sect:results}, evaluating accuracy.

At the level of the complete Swiss-Prot dataset (without organism stratification, the \emph{all} dataset), TFs and non-TFs exhibit highly similar length distributions. The median length is 281 amino acids for TFs and 286 amino acids for non-TFs, while the corresponding mean lengths are 331.6 and 319.9 amino acids, respectively. These results indicate that protein length is not a strongly discriminative feature in the global dataset.
Consistently, enforcing length balance leads only to a modest reduction in predictive performance. The mean accuracy of DeepTFactor decreases from 0.980 to 0.961, while GraphTFactor decreases from 0.987 to 0.980. These results suggest that both models primarily rely on features beyond sequence length, although a small contribution from length-related information cannot be excluded.

A clearer length-related signal emerges in organism-specific subsets. For the \emph{Eukaryotic} TFs, the lengths are substantially longer than those of non-TFs, with median lengths of 407 and 326 amino acids and mean lengths of 443.4 and 352.7 amino acids, respectively. This indicates a stronger potential for length-based discrimination in this group. Accordingly, DeepTFactor shows a more pronounced performance drop on the length-balanced dataset, decreasing from 0.963 to 0.917. This suggests that sequence length contributes more significantly to TF discrimination in eukaryotes, likely due to the interaction between biological length differences and the fixed-length encoding with zero-padding.

For the \emph{Prokaryotic} subset, TFs are generally shorter than non-TFs, with median lengths of 207 and 273 amino acids and mean lengths of 228.0 and 302.9 amino acids, respectively. Despite this clear shift in distribution, the impact of length balancing is limited: DeepTFactor decreases only slightly from 0.990 to 0.983, while GraphTFactor remains unchanged at 0.991. These results indicate that, in this subset, sequence length has only a marginal influence on predictive performance, particularly for GraphTFactor.

Finally, the \emph{Virus} subset shows a more pronounced effect of length balancing. Here, TFs are again shorter than non-TFs, with median lengths of 164 and 261 amino acids and mean lengths of 274.4 and 326.5 amino acids, respectively. In this case, both models experience a clearer performance decrease on the balanced dataset, with DeepTFactor dropping from 0.967 to 0.927 and GraphTFactor from 0.975 to 0.953. These results suggest that sequence length carries meaningful discriminative information in this subset and contributes to the predictive signal captured by both models.
Overall, these results indicate that protein sequence length plays a context-dependent role in transcription factor prediction. At the global level, length distributions between TFs and non-TFs are highly similar, and enforcing length balance only marginally affects predictive performance. This suggests that both DeepTFactor and GraphTFactor primarily rely on biologically meaningful sequence features rather than exploiting simple length-based cues.
However, a different picture emerges at the organism-specific level. In particular, eukaryotic proteins exhibit a clearer length separation between TFs and non-TFs, and this is accompanied by a more pronounced performance drop under length balancing, especially for DeepTFactor. This observation supports the hypothesis that, in settings where length is more strongly correlated with class identity, models leveraging fixed-length encodings may partially exploit length-related signals, including those introduced by zero-padding.
Conversely, prokaryotic subsets show heterogeneous behavior: in some cases, length balancing has a negligible impact, whereas in others, a measurable performance reduction is observed. This variability suggests that the contribution of sequence length is not intrinsic to the classification task itself, but rather depends on the underlying distributional properties of each dataset.
Importantly, GraphTFactor consistently exhibits smaller performance degradation than DeepTFactor, indicating greater robustness to length-induced biases. This is likely attributable to its architecture, which is less dependent on fixed-size representations and therefore less sensitive to padding artifacts.
In summary, while sequence length alone does not constitute a primary predictive feature for TF identification, it can serve as an informative proxy in specific organismal contexts and may introduce subtle biases in models that rely on fixed-length encodings. These findings highlight the importance of controlling for length effects when evaluating model robustness and underscore the advantages of architectures that are less sensitive to input length constraints.

\section{Ablation Study}
\label{sect:ablation}
To assess the contribution of each GraphTFactor component, we conducted five ablation experiments: 
\begin{enumerate*}[label=(\roman*)]
	\item ESM-2 embeddings with logistic regression (LR);
	\item ESM-2 embeddings with a multilayer perceptron (MLP);
	\item graph representations with a simple readout and LR;
	\item random amino acid embeddings in place of pretrained residue embeddings;
	\item randomized sequence segmentation in place of the proposed MDL-based segmentation.
\end{enumerate*}
An illustration of the complete experimental design was reported in Figure~\ref{fig:ablation} of the Appendix. 
Following the evaluation protocol adopted throughout this work, the analysis focuses on BA and Se, which are particularly informative for the TF prediction due to the class imbalance of the datasets employed in this study. For these experiments, we used the GraphTFactor variant based on the 30-layer ESM-2 model, which provides a practical trade-off between model capacity and computational cost. The results of these ablation experiments, along with a comparison with the whole GraphTFactor framework, are reported in Table~\ref{tab:ablation30}. All the results obtained with the other ESM-2 models can be found in Table~\ref{tab:ablation_all},Table~\ref{tab:ablation_euk}, Table~\ref{tab:ablation_prok}, and Table~\ref{tab:ablation_vir} of the Appendix. 
As shown in Table~\ref{tab:ablation30}, GraphTFactor achieves the highest balanced accuracy in the \emph{All}, \emph{Eukaryotic}, and \emph{Prokaryotic} datasets, and the highest sensitivity in the \emph{All} and \emph{Prokaryotic} datasets. The ESM-2 + MLP variant outperforms ESM-2 + LR in the All and Eukaryotic groups, indicating that a nonlinear classifier can make better use of the sequence embeddings in these subsets. In Prokaryotes, however, ESM-2 + LR performs better than ESM-2 + MLP. In Viruses, ESM-2 + MLP obtains the highest sensitivity, while its BA is close to that of GraphTFactor.
The graph readout with LR performs worse than the full model across all datasets, particularly in sensitivity, suggesting that the graph representation combined with a simple readout does not capture the benefits of the complete architecture. Replacing pretrained residue embeddings with random amino acid embeddings leads to the largest and most consistent performance reduction, with the lowest BA and Se in every group. This highlights the importance of pretrained sequence representations for GraphTFactor.
Randomized segmentation also reduces performance relative to MDL-based segmentation, most notably for Viruses, where BA decreases from 0.954 to 0.896 and Se from 0.880 to 0.777. The difference is smaller for Eukaryotes and Prokaryotes; in the latter, randomized segmentation remains competitive, although both metrics are lower than those of GraphTFactor. Overall, the results indicate that the contributions of the components vary across organism groups, with pretrained embeddings showing a consistent benefit and MDL-based segmentation showing its clearest advantage in the \emph{Virus} dataset.
Overall, the ablation study shows that GraphTFactor benefits from combining its components, although their individual contributions vary across datasets. The 30-layer ESM-2 model provides a practical balance between representational capacity and computational cost for these experiments. Section~\ref{sect:ablationsupp} of the Appendix provides a more detailed description of the ablation experiments and results obtained.

\begin{table}[]
	\centering
	\caption{Ablation Study results for the GraphTFactor variant using the ESM-2 model. Results are reported as Balanced Accuracy (BA) and Sensitivity (Se). Best results are in bold, and the second-best are underlined.}
	\label{tab:ablation30}
	\begin{adjustbox}{max totalsize={\textwidth}{0.85\textheight},center}
		\begin{tabular}{@{}lcc@{}}
			\toprule
			\multicolumn{3}{c}{\emph{All}}                                         \\ \midrule
			Experiment                    & BA & Se \\ \midrule
			GraphTFactor                  & $\pmb{0.973 \pm 0.002}$                  &   $\pmb{0.955 \pm 0.004}$          \\
			ESM-2 + LR                    &  $0.946 \pm 0.003$                 &   $0.913 \pm 0.001$          \\
			ESM-2 + MLP                   &  $0.961 \pm 0.004$                 &    $0.934 \pm 0.001$         \\ 
			Graph Readout + LR            & $0.868 \pm 0.006$                  &  $0.782 \pm 0.011$           \\
			GraphTF with random AA emb.   & $0.826 \pm 0.018$                   &  $0.675 \pm 0.044$           \\
			GraphTF with random seq. seg. &      $\underline{0.955 \pm 0.003}$             &  $\underline{0.926 \pm 0.009}$        \\ \bottomrule
			
			\multicolumn{3}{c}{\emph{Eukaryotic}}                                         \\ \midrule
			Experiment                    & BA & Se \\ \midrule
			GraphTFactor                  & $\pmb{0.959 \pm 0.005}$                   & $\pmb{0.931 \pm 0.012}$             \\
			ESM-2 + LR                    &    $0.934 \pm 0.006$               & $0.889 \pm 0.013$             \\
			ESM-2 + MLP                   & $\underline{0.942 \pm 0.011}$                  &  $\underline{0.898 \pm 0.025}$           \\ 
			Graph Readout + LR            &      $0.876 \pm 0.006$             & $0.798 \pm 0.015$            \\
			GraphTF with random AA emb.   &   $0.806 \pm 0.009$                & $0.632 \pm 0.021$             \\
			GraphTF with random seq. seg. &    $0.932 \pm 0.008$               &   $0.884 \pm 0.020$       \\ \bottomrule

			\multicolumn{3}{c}{\emph{Prokaryotic}}                                         \\ \midrule
			Experiment                    & BA & Se \\ \midrule
			GraphTFactor                  & $\pmb{0.993 \pm 0.001}$                  &  $\pmb{0.984 \pm 0.005}$           \\
			ESM-2 + LR                    &    $0.976 \pm 0.002$               &      $\underline{0.965 \pm 0.005}$       \\
			ESM-2 + MLP                   &      $0.954 \pm 0.013$             & $0.912 \pm 0.027$            \\ 
			Graph Readout + LR            &     $0.936 \pm 0.003$              & $0.896 \pm 0.005$            \\
			GraphTF with random AA emb.   &    $0.894 \pm 0.006 $               &  $0.840 \pm 0.019$           \\
			GraphTF with random seq. seg. &    $\underline{0.979 \pm 0.002}$               & $\underline{0.965 \pm 0.005}$          \\ \bottomrule

			\multicolumn{3}{c}{\emph{Virus}}                                         \\ \midrule
			Experiment                    & BA & Se \\ \midrule
			GraphTFactor                  &    $\pmb{0.954 \pm 0.010}$               & $\underline{0.880 \pm 0.049}$            \\
			ESM-2 + LR                    &     $0.841 \pm 0.034$              & $0.719 \pm 0.067$             \\
			ESM-2 + MLP                   &     $\underline{0.951 \pm 0.014}$              &   $\pmb{0.936 \pm 0.021}$          \\ 
			Graph Readout + LR            &   $0.879 \pm 0.019$                & $0.825 \pm 0.041$            \\
			GraphTF with random AA emb.   &    $0.828 \pm 0.035$               & $0.739 \pm 0.082$            \\
			GraphTF with random seq. seg. &   $0.896 \pm 0.025$                & $0.777 \pm 0.064$          \\ \bottomrule
		\end{tabular}
	\end{adjustbox}
\end{table}

\section{Discussions}
\label{sect:discussion}
The experimental activity demonstrates that GraphTFactor is a competitive approach for TF prediction, achieving performance comparable to, and in several settings better than, DeepTFactor.
A first observation concerns the segmentation of protein sequences via MDL. Unlike fixed-length k-mer segmentation used in our previous work \cite{amato2024proteins}, the MDL-based procedure partitions each sequence into variable-length subsequences without requiring the choice of an additional hyperparameter such as $k$. The ablation study confirms that this segmentation consistently outperforms a length-matched random partitioning strategy, although the margin is modest and narrows further as larger ESM-2 models are used. This indicates that meaningful sequence partitioning contributes positively, but is not the primary source of GraphTFactor's predictive power. As a consequence, GraphTFactor can capture local sequence patterns more flexibly while preserving biologically relevant information that might otherwise be fragmented or lost under rigid fixed-length decompositions. 
A second, larger contribution comes from the ESM-2 backbone. Across nearly all datasets, performance improves as the depth and representational capacity of ESM-2 increase, and the ablation shows that replacing pretrained embeddings with random vectors causes by far the largest performance drop observed in this study (e.g., BA from ~0.97–0.98 to ~0.82 on the \emph{All} dataset). This indicates that the biological information encoded by the pretrained protein language model, rather than the graph topology itself, is the main driver of GraphTFactor's predictive performance. The graph construction and the GraphSAGE message-passing stage provide a consistent but secondary improvement over classifiers operating on global ESM-2 embeddings alone, particularly on the \emph{All} and \emph{Eukaryotic} datasets. An important exception to this pattern is the Virus dataset. Here, a simple MLP or logistic-regression classifier trained on mean-pooled ESM-2 embeddings matches or slightly exceeds GraphTFactor when the largest ESM-2 variants are used. This suggests that, for this subset, global sequence-level information is largely sufficient for accurate classification, and the additional regional structure introduced by the graph does not provide a consistent advantage. Given the small size of the \emph{Virus} dataset (538 TF sequences) and the correspondingly high variance across folds, this result should be interpreted with some caution, but it is a genuine and informative negative result rather than an artifact to be explained away: it indicates that the benefit of graph-based modeling is dataset-dependent rather than universal. 
A practical advantage of GraphTFactor, independent of the classification-performance comparison, is its ability to process sequences longer than 1000 amino acids without truncation, a constraint inherent to DeepTFactor's fixed-length one-hot encoding. The length-balancing experiments in Section 4 further show that GraphTFactor's performance degrades less than DeepTFactor's when sequence length is controlled for as a confound, particularly in the \emph{Eukaryotic} subset (0.963 $\rightarrow$ 0.917 for DeepTFactor vs. 0.972 $\rightarrow$ 0.967 for GraphTFactor). This is consistent with the graph-based representation being less reliant on padding-related artifacts. Taken together, these results support a more circumscribed conclusion than "GraphTFactor consistently outperforms DeepTFactor across diverse biological domains": the framework provides a measurable and reasonably consistent improvement, driven primarily by the quality of pretrained protein embeddings and secondarily by graph-based relational modeling, with the graph component's contribution being smallest — and occasionally negative relative to simpler baselines — on the most heterogeneous and data-scarce subset considered. Some limitations should be acknowledged. Larger ESM-2 models yield better predictive performance but require substantially greater computational resources for feature extraction; intermediate configurations (30–33 layers) likely offer the best accuracy/cost trade-off. Once trained, inference for a new protein requires only graph construction and a forward pass through the trained GNN, so deployment cost is substantially lower than training cost.
In conclusion, the results show that GraphTFactor is a flexible, scalable framework that adapts effectively to datasets with diverse biological characteristics. Its consistent performance improvements over DeepTFactor suggest that graph-based learning constitutes a promising direction for future research in transcription factor prediction and, more broadly, in protein function annotation.

\section{Conclusions}
\label{sect:concl}
In this work, we presented GraphTFactor, a graph-based framework for TF prediction that combines MDL-based adaptive sequence segmentation, ESM-2 protein language model embeddings, and a GraphSAGE-based GNN classifier. Across four organism-specific, heterogeneous datasets derived from the Swiss-Prot database, GraphTFactor achieved performance comparable to, and in most settings better than, DeepTFactor, with the largest gains on the all and Eukaryotic datasets.
The ablation study indicates that this improvement stems primarily from the pretrained ESM-2 embeddings, with the graph construction and message-passing stages providing a smaller, complementary contribution. On the \emph{Virus} dataset, this contribution was not always positive relative to simpler classifiers operating on global embeddings, suggesting that the value of graph-based modeling depends on the structural and evolutionary heterogeneity of the target class. The framework also showed reduced sensitivity to sequence-length-related biases compared to a fixed-length encoding baseline, and it removes the need to truncate or pad proteins longer than 1000 amino acids — a practical advantage independent of raw classification accuracy. To support reproducibility, we release GraphTFactor as an open-source software package together with a web application for TF prediction on custom sequences. Future work will focus on (i) applying explainability methods such as GECo \cite{amato2026geco} to identify the specific subsequences driving individual predictions — a capability we motivate but do not demonstrate in this study, and (ii) extending GraphTFactor from protein-level to residue-level prediction (e.g., DNA-binding site localization). 

\section*{Acknowledgements}
All the authors acknowledge the support of the INdAM-GNCS Project 2026 "Graph, Generative, and state-space models for Multi-Omics integrative Discovery" CUP E53C25002010001, the INdAM-GNCS project "M-GIST - Medical Graph-Integrated Segmentation Techniques for Classification," CUP E53C25002010001 and the INdAM-GNCS project 2026 "Metodi Numerici e Machine Learning per Sistemi Dinamici Complessi: Identificazione, Approssimazione", CUP E53C25002010001.

\section*{Code and Data Availability}
The datasets generated for this study and the source code are available in the following GitHub repository~\cite{amato2026graphtfactorgit}. A dedicated web application is available at~\cite{amato2026graphtfactorweb}, and the Python package developed for this study is available at~\cite{amato2026graphtfactorpack}.

\section*{Conflict of Interest}
The authors declare no conflict of interest.

\begin{appendix}
		\section{Datasets generation}
	\label{sect:suppdata}
\begin{figure}[ht]
		\centering
		\includegraphics[width=1\linewidth]{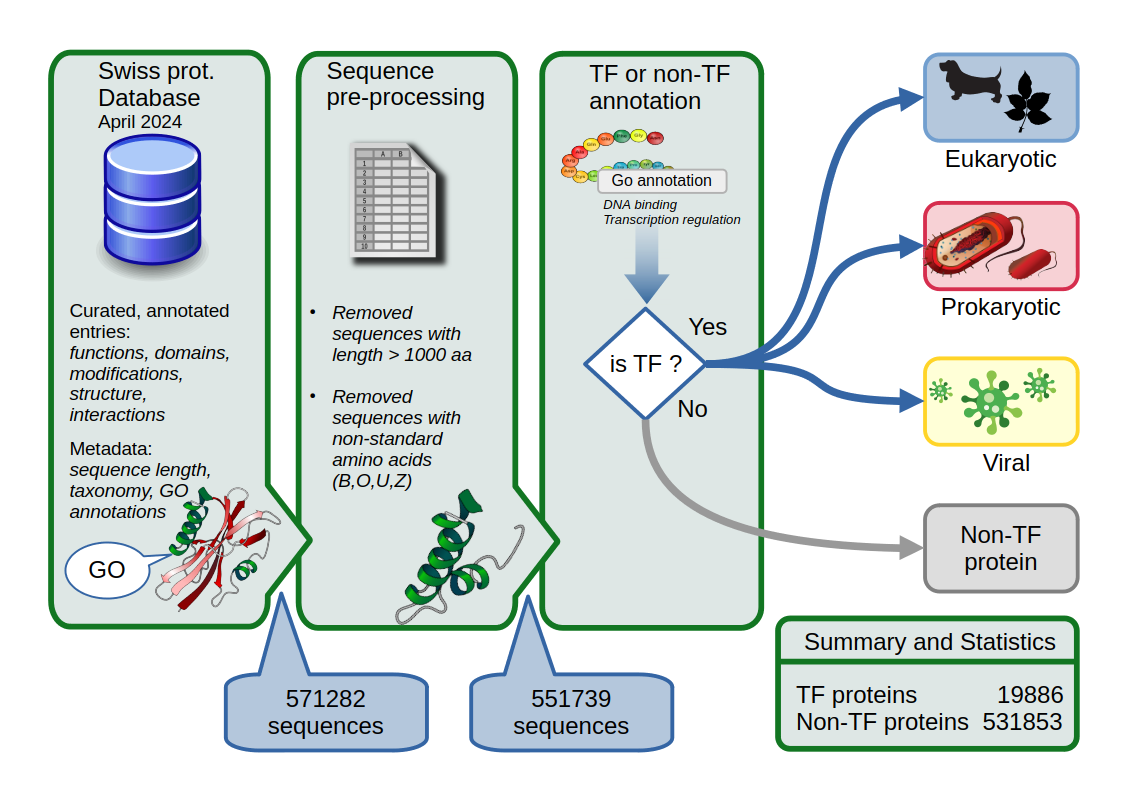}
		\caption{An illustration of the procedure used for the protein sequence TF annotation and organism identification.}
		\label{fig:annot}
	\end{figure}

\begin{figure}[ht]
		\centering
		\includegraphics[width=0.75\linewidth]{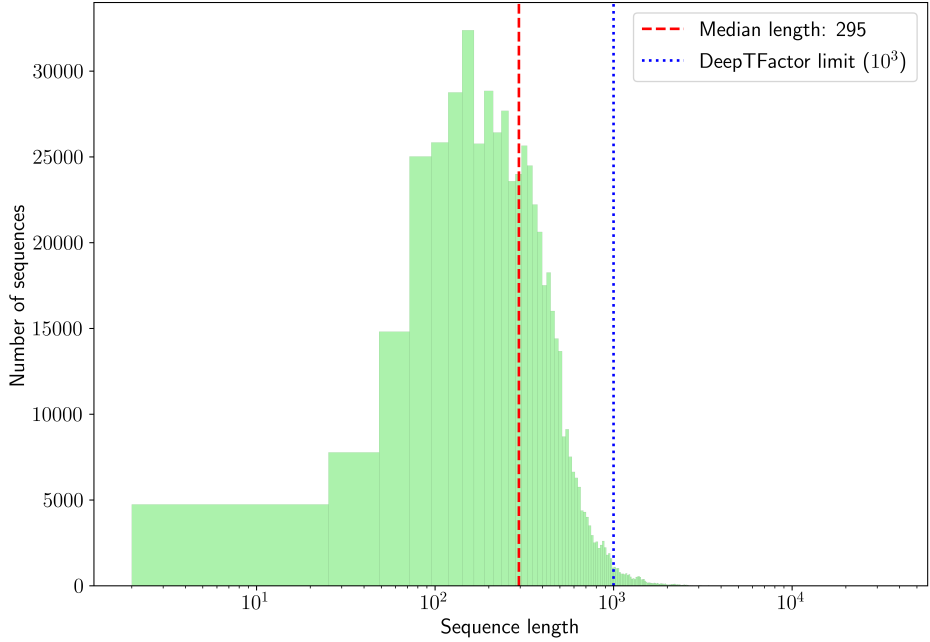}
		\caption{Histogram showing the distribution of sequence lengths. The red line represents the median value, equal to 295}\label{fig:lendist}
	\end{figure}

	To conduct our experimental activity, we have constructed four different datasets using the SwissProt database downloaded in April 2024\footnote{\url{https://www.uniprot.org/help/downloads}}. SwissProt is a curated database of protein sequences that provides high-quality, manually annotated information. Each entry includes functional descriptions, such as biological roles and molecular activities, as well as details on protein domains, post-translational modifications, and links to experimentally validated data. Swiss-Prot also incorporates information on three-dimensional structures, protein–protein interactions, subcellular localization, and cross-references to numerous external biological databases. In the Swiss-Prot version considered in this work, the dataset contains 571282 protein sequences. To establish whether a protein is a Transcription Factor (TF) or not, we followed the same approach used in \cite{kim2021deeptfactor}. In particular, this annotation protocol relies on the Gene Ontology (GO) terms. A GO  annotation provides structured information about a protein’s molecular function, biological process, and cellular component, using a controlled vocabulary. Each GO term is assigned through experimental evidence or high-quality manual curation. For the annotation of a protein as a TF, we have used the following criteria:
	\begin{enumerate}
		\item has a GO annotation for the TF activity;
		\item has both a DNA binding-related GO annotation and a transcription regulation-related GO annotation.
	\end{enumerate}
	Furthermore, to perform a more in-depth analysis, we have categorized them based on their organism type: eukaryotic, prokaryotic, or viral. For this categorization, we defined the following taxonomic terms:
	\begin{enumerate}
		\item eukaryotic terms: \textit{Eukaryota};
		\item prokaryotic terms: \textit{Prokaryota} and \textit{Archaea};
		\item viral terms: \textit{Virus}.
	\end{enumerate}
	An illustration depicting the annotation procedure is reported in Figure~\ref{fig:annot}.
	The Swiss-Prot database contains proteins with a minimum sequence length of two amino acids and a maximum length of 35213 amino acids, a median length of 295, and a mean length of about 362 amino acids. In Figure~\ref{fig:lendist}, we report the histogram of sequence lengths, and it is possible to notice that the majority of sequences (about 96\%) have lengths in the range $[2, 1000]$; hence, we have decided to filter out the sequences with more than 1000 amino acids (i.e., the limit imposed by DeepTFactor), not considering further those that contain no standard amino acids (B, O, U, Z). After this filtering operation, we found 19986 TF protein sequences and 531853 non-TF proteins. 
	
	Since the dataset is highly imbalanced, we have discarded many non-TF proteins. After determining whether a protein contains a transcription factor and finding the species type, we generate four datasets:
	\begin{itemize}
		\item \textit{All dataset}: containing 19986 TF sequences and 59958 no-TF ones;
		\item \textit{Eukaryotic dataset}: containing 9494 TF sequences and 28482 no-TF ones
		\item \textit{Prokaryotic dataset}: containing 9954 TF sequences and 29862 no-TF ones.
		\item \textit{Virus dataset}: containing 538 TF sequences and 1614  no-TF ones.
	\end{itemize}
	It is essential to note that we have selected the proportion of TF and no-TF sequences using the same criteria reported in \cite{kim2021deeptfactor,amato2024proteins}.

	\section{Optimal segmentation via dynamic programming}
	\label{sect:suppdp}
\begin{figure}[h]
		\centering
		\includegraphics[width=1\linewidth]{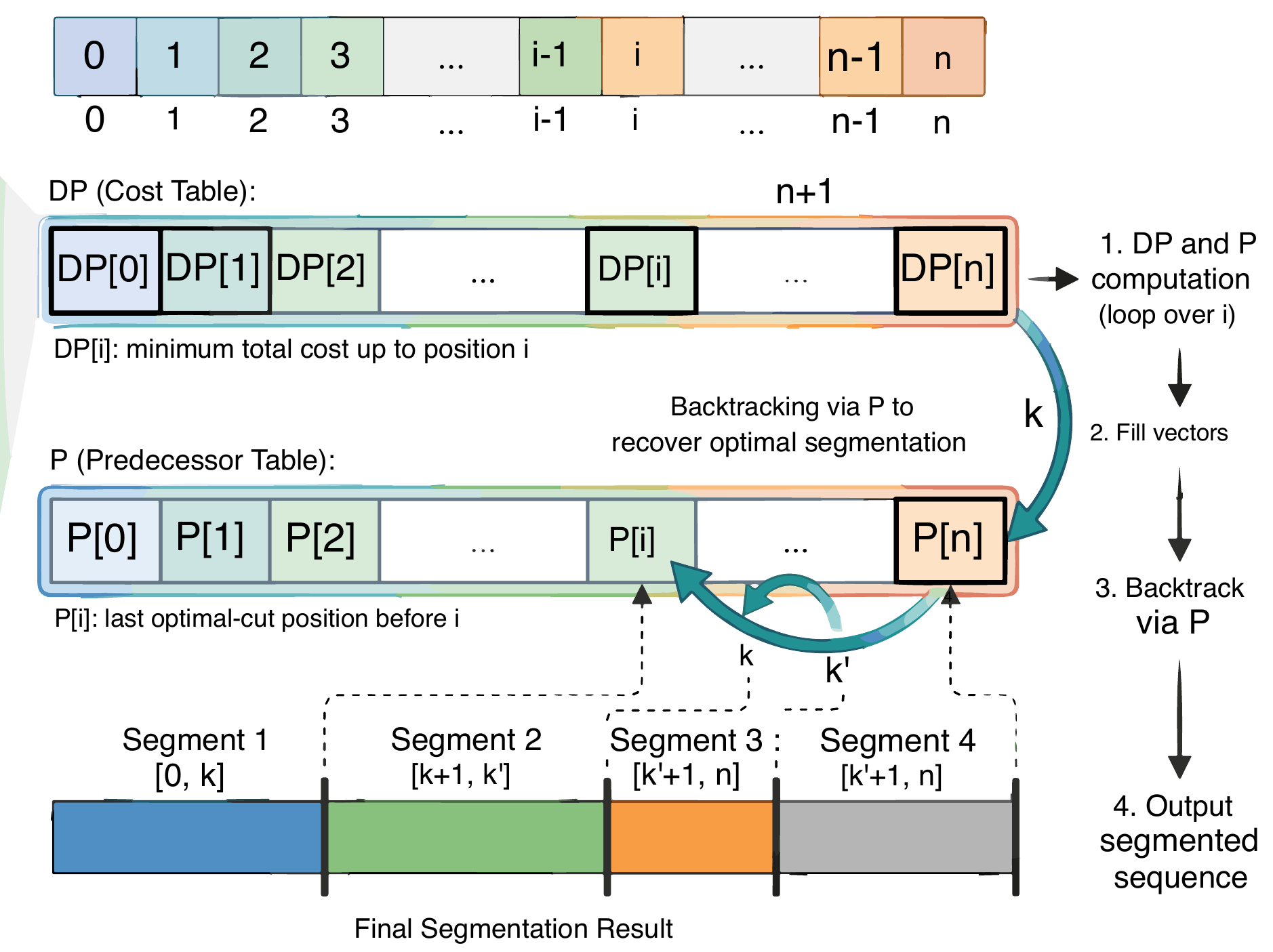}
		\caption{An illustration of the dynamic programming strategy employed.}
		\label{fig:dp}
	\end{figure}

	To obtain the optimal segmentation, we employed a dynamic programming (DP) approach \cite{bellman1957dynamic}. In particular, we define two vectors of length $n+1$ (where $n$ is the sequence's length), denoted as $DP$ and $P$. $DP[i]$ is used to store the minimum total cost for segmenting the prefix of the sequence up to position $i$, while $P[i]$ records the last segmentation point that yields the minimum cost. The DP recursion is defined as:
	\begin{equation}
		\label{eq:dprec}
		DP[i] = \min_{0 \leq j \leq i} \left\{ DP[j] + \mathcal{C}(S_{j+1:i}) + \log{n} \right\},
	\end{equation}
	where $DP[i]$ represents the minimum total cost of segmenting the prefix of the sequence up to position $i$,  and $DP[j]$ represents the minimum total cost of segmenting the prefix up to position $j$, while $\mathcal{C}(S_{j+1:i})$ is the encoding cost of the segment $S_{j+1:i}$.
	This recursion enables the algorithm to compute the minimum cumulative cost for every prefix of the sequence while storing the information required to reconstruct the optimal segmentation. Let $P[i]$ denote the predecessor index associated with the optimal segmentation of the prefix ending at position $i$:
	\begin{equation}
		P[i] = \argmin_{0 \leq j \leq i}
		\left\{
		DP[j] + \mathcal{C}(S_{j+1:i}) + \log{n}
		\right\}.
	\end{equation}
	where the $\argmin(\cdot)$ operator returns the index $j \in \{0,\ldots,i\}$ that minimizes the objective function. Consequently, $P[i]$ stores the position of the last optimal cut preceding residue $i$. After the dynamic programming table has been filled, the optimal segmentation is recovered by backtracking through the predecessor array $P$. Starting from the last sequence position, the algorithm repeatedly retrieves the previous cut stored in $P$, extracts the corresponding segment, and updates the current position to the retrieved index. The procedure terminates when the beginning of the sequence is reached, and the collected segments are reversed to restore their original order.
	This dynamic programming strategy yields an optimal variable-length segmentation, allowing each segment to adapt to the underlying sequence characteristics while avoiding unnecessary fragmentation through the regularization term $\lambda$.
	An illustration of the dynamic programming process employed is reported in Figure~\ref {fig:dp}.
	
		\section{Results}
	\label{sect:suppresults}
	Table~\ref{table:results} reports a comparative evaluation of DeepTFactor~\cite{kim2021deeptfactor} and GraphTFactor across the complete evaluation set and three taxonomic subsets: Eukaryotic, Prokaryotic, and Virus. For GraphTFactor, the analysis considers residue representations extracted from different layers of the ESM-2 model (layers 6, 12, 30, 33, 36, and 48), allowing the effect of the representation depth on transcription factor prediction performance to be examined.
	
	Performance is reported using Accuracy (Acc), Balanced Accuracy (BA), Sensitivity (Se), Specificity (Spe), F1-Score (F1), and area under the receiver operating characteristic curve (AUC). Accuracy provides an overall measure of correct predictions, whereas balanced accuracy accounts for the contribution of both classes and is therefore particularly informative in the presence of class imbalance. Sensitivity and specificity quantify the ability to correctly identify transcription factors and non-transcription factors, respectively, while the F1-score summarizes the balance between precision and sensitivity. AUC provides a threshold-independent measure of the discriminative ability of the models.
	
	For each dataset and evaluation metric, the table highlights the three highest mean performance values across the considered methods using distinct cell colors, with the best, second-best, and third-best results denoted accordingly. The reported values correspond to the mean performance and standard deviation obtained across the experimental repetitions. This organization provides a direct comparison between the original DeepTFactor model and the different GraphTFactor configurations, while also allowing potential differences in performance across taxonomic groups to be assessed.
	
	\begin{landscape}
		\begin{table*}
			\centering
			\caption{Comparison of DeepTFactor \cite{kim2021deeptfactor} and GraphTFactor across the evaluated datasets. Results are reported in terms of accuracy (Acc), balanced accuracy (BA), sensitivity (Se), specificity (Spe), F1-score (F1), and area under the ROC curve (AUC). Cell colors indicate the ranking of the results for each dataset and metric: \colorbox{best}{best}, \colorbox{second}{second-best}, and \colorbox{third}{third-best}.}
			\label{table:results}
			\resizebox{\linewidth}{!}{
				\begin{tabular}{llllllll}
					\toprule\toprule
					\textbf{Dataset} & \textbf{Method} & \textbf{Acc} & \textbf{BA} & \textbf{Se} & \textbf{Spe} & \textbf{F1} & \textbf{AUC} \\
					\midrule
					
					\multirow{7}{*}{\textit{All}}
					& DeepTFactor     & $0.980 \pm 0.001$ & $0.970 \pm 0.001$ & $0.949 \pm 0.003$ & \cellcolor{third}$0.990 \pm 0.003$ & $0.959 \pm 0.003$ & $0.990 \pm 0.001$ \\
					& GraphTFactor 6  & $0.976 \pm 0.002$  & $0.966 \pm 0.002$ & $0.946 \pm 0.005$ & $0.986 \pm 0.003$  & $0.951 \pm 0.003$  & $0.993 \pm 0.006$  \\
					& GraphTFactor 12 & $0.979 \pm 0.002$  & $0.970 \pm 0.002$  & $0.953 \pm 0.003$  & $0.988 \pm 0.003$  & $0.958 \pm 0.004$  & \cellcolor{third}$0.994 \pm 0.000$  \\
					& GraphTFactor 30 & $0.982 \pm 0.001$  & $0.973 \pm 0.002$  & $0.955 \pm 0.004$  & \cellcolor{best}$0.991 \pm 0.001$ & $0.964 \pm 0.002$ & \cellcolor{second}$0.995 \pm 0.000$  \\
					& GraphTFactor 33 & \cellcolor{third}$0.985 \pm 0.001$  & \cellcolor{third}$0.979 \pm 0.001$  &  \cellcolor{third}$0.966 \pm 0.004$ & \cellcolor{best}$0.991 \pm 0.001$  & \cellcolor{third}$0.971 \pm 0.001$  & \cellcolor{best}$0.996 \pm 0.000$  \\
					& GraphTFactor 36 & \cellcolor{second}$0.987 \pm 0.001$  & \cellcolor{second}$0.982 \pm 0.001$  & \cellcolor{second}$0.973 \pm 0.003$ & \cellcolor{second}$0.991 \pm 0.002$  & \cellcolor{second}$0.974 \pm 0.003$  & \cellcolor{best}$0.996 \pm 0.000$  \\
					& GraphTFactor 48 & \cellcolor{best}$0.987 \pm 0.000$ & \cellcolor{best}$0.983 \pm 0.001$  & \cellcolor{best}$0.974 \pm 0.004$  & \cellcolor{best}$0.991 \pm 0.001$  & \cellcolor{best}$0.974 \pm 0.001$  & \cellcolor{best}$0.996 \pm 0.000$  \\
					
					\midrule\midrule
					
					\multirow{7}{*}{\textit{Eukaryotic}}
					& DeepTFactor & $0.963 \pm 0.006$  & $0.948 \pm 0.09$  & $0.919 \pm 0.026$  & $0.978 \pm 0.012$  & $0.926 \pm 0.011$  & $0.982 \pm 0.002$  \\
					& GraphTFactor 6  &$0.961 \pm 0.002$  & $0.946 \pm 0.006$  & $0.902 \pm 0.017$  & $0.980 \pm 0.005$  & $0.920 \pm 0.006$  & $0.986 \pm 0.016$  \\
					& GraphTFactor 12 & $0.967 \pm 0.003$  & $0.955 \pm 0.005$ & $0.931 \pm 0.009$  & $0.979 \pm 0.003$  & $0.934 \pm 0.007$  & $0.987 \pm 0.001$ \\
					& GraphTFactor 30 & $0.972 \pm 0.002$  & $0.959 \pm 0.005$  & $0.931 \pm 0.012$  & \cellcolor{second}$0.988 \pm 0.003$  & $0.944 \pm 0.005$  & \cellcolor{third}$0.990 \pm 0.001$  \\
					& GraphTFactor 33 & \cellcolor{third}$0.976 \pm 0.002$  & \cellcolor{third}$0.963 \pm 0.004$  & \cellcolor{third}$0.938 \pm 0.008$  & \cellcolor{best}$0.988 \pm 0.002$  & \cellcolor{third}$0.951 \pm 0.005$  & \cellcolor{best}$0.992 \pm 0.001$  \\
					& GraphTFactor 36 & \cellcolor{second}$0.978 \pm 0.002$  & \cellcolor{second}$0.968 \pm 0.005$  & \cellcolor{second}$0.949 \pm 0.010$  & \cellcolor{best}$0.988 \pm 0.002$   & \cellcolor{second}$0.956 \pm 0.005$  & \cellcolor{best}$0.992 \pm 0.001$  \\
					& GraphTFactor 48 & \cellcolor{best}$0.979 \pm 0.002$  & \cellcolor{best}$0.971 \pm 0.003$  & \cellcolor{best}$0.955 \pm 0.007$  & \cellcolor{third}$0.986 \pm 0.002$  & \cellcolor{best}$0.957 \pm 0.004$  & \cellcolor{second}$0.991 \pm 0.001$  \\
					
					\midrule\midrule
					
					\multirow{7}{*}{\textit{Prokaryotic}}
					& DeepTFactor &$0.990 \pm 0.001$  & $0.984 \pm 0.001$   & $0.972 \pm 0.003$  & $0.996 \pm 0.001$  & $0.980 \pm 0.002$  & $0.994 \pm 0.001$  \\
					& GraphTFactor 6  & $0.988 \pm 0.001$  & $0.983 \pm 0.007$  & $0.974 \pm 0.006$  & $0.993 \pm 0.001$  & $0.982 \pm 0.003$  & \cellcolor{third}$0.997 \pm 0.000$  \\
					& GraphTFactor 12 & $0.991 \pm 0.001$  & $0.987 \pm 0.002$  & $0.980 \pm 0.004$  & $0.995 \pm 0.001$ & $0.977 \pm 0.002$  & $0.997 \pm 0.001$  \\
					& GraphTFactor 30 & \cellcolor{third}$0.993 \pm 0.001$ & $0.990 \pm 0.001$  & $0.984 \pm 0.005$  & \cellcolor{third}$0.996 \pm 0.000$  & $0.986 \pm 0.002$  & \cellcolor{second}$0.998 \pm 0.001$  \\
					& GraphTFactor 33 & \cellcolor{second}$0.994 \pm 0.000$  & \cellcolor{third}$0.991 \pm 0.002$  & \cellcolor{third}$0.986 \pm 0.003$  & \cellcolor{third}$0.966 \pm 0.000$  & \cellcolor{third}$0.988 \pm 0.001$  & \cellcolor{second}$0.998 \pm 0.001$  \\
					& GraphTFactor 36 & \cellcolor{best}$0.995 \pm 0.001$  & \cellcolor{best}$0.993 \pm 0.001$  & \cellcolor{best}$0.989 \pm 0.003$  & \cellcolor{second}$0.997 \pm 0.001$  & \cellcolor{best}$0.990 \pm 0.002$  & \cellcolor{best}$0.998 \pm 0.000$  \\
					& GraphTFactor 48 & \cellcolor{second}$0.994 \pm 0.000$ & \cellcolor{second}$0.992 \pm 0.001$  & \cellcolor{second}$0.987 \pm 0.002$  & \cellcolor{best}$0.997 \pm 0.000$  & \cellcolor{second}$0.989 \pm 0.001$  & \cellcolor{best}$0.998 \pm 0.000$  \\
					
					\midrule\midrule
					
					\multirow{7}{*}{\textit{Virus}}
					& DeepTFactor & \cellcolor{second}$0.967 \pm 0.009$  & $0.937 \pm 0.019$  & $0.879 \pm 0.038$    &  \cellcolor{best}$0.996 \pm 0.003$  & $0.929 \pm 0.021$  & $0.970 \pm 0.015$  \\
					& GraphTFactor 6  & $0.931 \pm 0.012$  & $0.908 \pm 0.023$  & $0.862 \pm 0.059$  & $0.954 \pm 0.018$  & $0.862 \pm 0.059$  & $0.970 \pm 0.010$  \\
					& GraphTFactor 12 & $0.953 \pm 0.010$  & $0.929 \pm 0.023$  & $0.881 \pm 0.053$  & $0.978 \pm 0.022$  & $0.904 \pm 0.024$  & $0.979 \pm 0.009$  \\
					& GraphTFactor 30 & $0.954 \pm 0.010$ & $0.929 \pm 0.022$  & $0.880 \pm 0.049$  & $0.978 \pm 0.008$  & $0.905 \pm 0.023$  & $0.975 \pm 0.012$  \\
					& GraphTFactor 33 & \cellcolor{second}$0.967 \pm 0.009$  & \cellcolor{second}$0.948 \pm 0.016$  & \cellcolor{third}$0.908 \pm 0.034$  & \cellcolor{third}$0.987 \pm 0.008$ & \cellcolor{second}$0.933 \pm 0.020$  & \cellcolor{best}$0.990 \pm 0.004$  \\
					& GraphTFactor 36 & \cellcolor{third}$0.966 \pm 0.018$ & \cellcolor{third}$0.948 \pm 0.024$  & \cellcolor{best}$0.914 \pm 0.037$  & $0.983 \pm 0.015$  & \cellcolor{third}$0.930 \pm 0.037$  & \cellcolor{second}$0.985 \pm 0.010$  \\
					& GraphTFactor 48 & \cellcolor{best}$0.971\pm 0.007$ & \cellcolor{best}$0.951 \pm 0.010$  & \cellcolor{second}$0.912 \pm 0.021$  & \cellcolor{second}$0.990 \pm 0.007$  & \cellcolor{best}$0.940 \pm 0.015$  & \cellcolor{third}$0.981 \pm 0.009$  \\
					
					\bottomrule\bottomrule
				\end{tabular}
			}
		\end{table*}
	\end{landscape}
	
	\section{Ablation Study}
	\label{sect:ablationsupp}
\begin{figure}
		\centering
		\begin{subfigure}[b]{0.99\textwidth}
			\centering
			\includegraphics[width=\textwidth]%
			{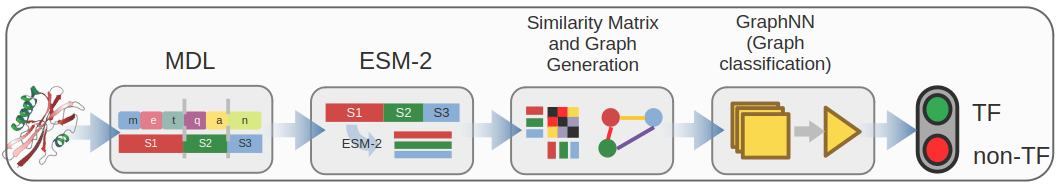}
			\caption{The proposed pipeline.}
			\label{fig:proposed_pipeline}
		\end{subfigure}
		\hfill
		
		\begin{subfigure}[b]{0.99\textwidth}
			\centering
			\includegraphics[width=\textwidth]%
			{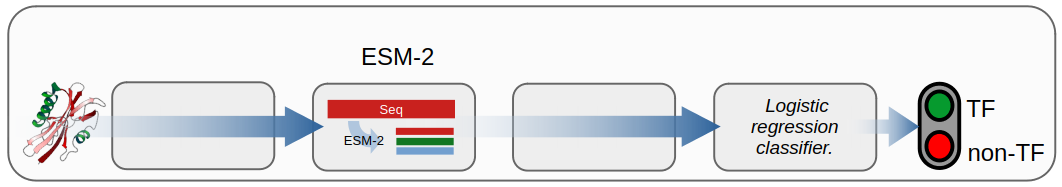}
			\caption{The baseline for comparison.}
			\label{fig:the_baseline}
		\end{subfigure}
		\hfill
		
		\begin{subfigure}[b]{0.99\textwidth}
			\centering
			\includegraphics[width=\textwidth]%
			{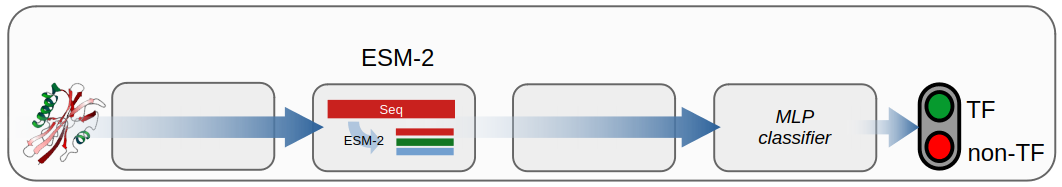}
			\caption{Exclusion of the graph representation and GNN.}
			\label{fig:excluding_graph_GNN}
		\end{subfigure}
		\hfill
		
		\begin{subfigure}[b]{0.99\textwidth}
			\centering
			\includegraphics[width=\textwidth]%
			{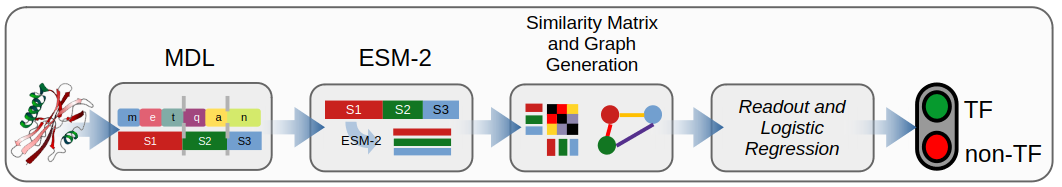}
			\caption{Exclusion of the Graph Neural Network.}
			\label{fig:excluding_GNN}
		\end{subfigure}
		\hfill
		
		\begin{subfigure}[b]{0.99\textwidth}
			\centering
			\includegraphics[width=\textwidth]%
			{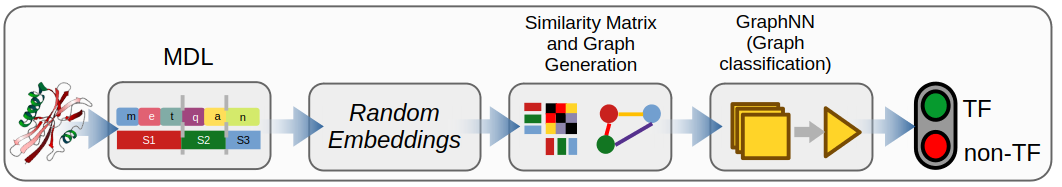}
			\caption{Exclusion of the ESM-2 embedding.}
			\label{fig:excluding_EMS-2}
		\end{subfigure}
		\hfill
		
		\begin{subfigure}[b]{0.99\textwidth}
			\centering
			\includegraphics[width=\textwidth]%
			{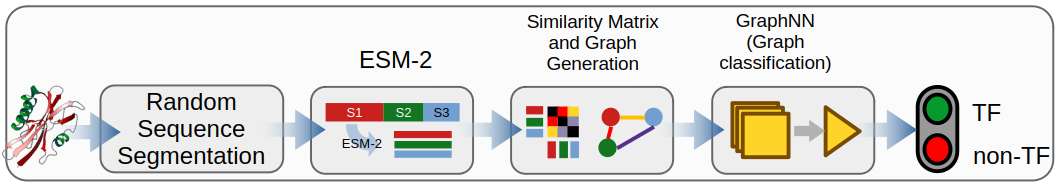}
			\caption{Exclusion of the MDL segmentation of the sequences.}
			\label{fig:excluding_MDL}
		\end{subfigure}
		\hfill   
		
		\caption{A representation schema of the proposed ablation study.}
		\label{fig:ablation}
	\end{figure}

	To better understand the individual contribution of each GraphTFactor component, we conducted a comprehensive ablation study to evaluate the impact of the protein language model, the graph construction procedure, the graph neural network, and the proposed MDL-based sequence segmentation. 
	Four complementary experiments were conducted: 
	\begin{enumerate*}[label=(\roman*)]
		\item a baseline considering ESM-2 embeddings and logistic regression,
		\item removing the graph and considering the ESM-2 embeddings and the MLP classifier,
		\item graph representation with simple readout and logistic regression,
		\item replacing pretrained residue embeddings with random amino acid embeddings, and
		\item substituting the proposed MDL segmentation with a randomized segmentation strategy.
	\end{enumerate*}
	The complete experimental design of the ablation study is illustrated in Figure~\ref{fig:ablation}.

	Following the evaluation protocol adopted throughout this work, the analysis focuses on Balanced Accuracy (BA) and Sensitivity (Se), which are particularly informative for the TF prediction due to the class imbalance of the datasets employed in this study. All the detailed results supporting the considerations made in the following subsections are reported in Tables~\ref{tab:ablation_all}, \ref{tab:ablation_euk}, \ref{tab:ablation_prok} and \ref{tab:ablation_vir} of the Appendix.

	\subsection{The baseline: ESM-2 Embeddings with Logistic Regression}
	This experiment investigates whether the performance of GraphTFactor derives from the expressive embedding representations learned by ESM-2 or from the proposed graph-based neural architecture. For this experiment, proteins are encoded by mean-pooled ESM-2 embeddings and subsequently classified using a Logistic regression classifier, providing a strong linear baseline (see Figure~\ref{fig:the_baseline}). The conducted experiments allow us to demonstrate that the ESM-2 representations already contain discriminative information, with performance steadily improving as larger language models are employed. For the \emph{all} dataset, BA increases from $0.904$ using the 6-layer ESM model to $0.980$ using the 48-layer model.  Nevertheless, GraphTFactor consistently achieves superior performance across almost all embedding sizes, improving BA up to 0.966–0.983 while simultaneously increasing Se from 0.840–0.970 to 0.946–0.974. The largest improvement is more evident for the smallest ESM-2 models, indicating that graph encoding and subsequent GNN classification can compensate for less informative sequence embeddings. As the embedding quality improves, the performance gap gradually decreases, suggesting that increasingly expressive language models already encode richer contextual information. However, even for the largest ESM variants, GraphTFactor maintains a measurable advantage on the complete dataset, demonstrating that graph construction and message passing capture complementary structural information beyond global protein representations. A similar trend is observable for the \emph{Eukaryotic dataset}, where GraphTFactor outperforms the Logistic Regression classifier across all the ESM-2 variants. For the \emph{Prokaryotic} dataset, both approaches achieve high performance, with BA approaching 0.999, indicating that transcription factor prediction in prokaryotes is comparatively easier. In this setting, the improvement introduced by GraphTFactor becomes naturally smaller because the baseline is already close to saturation. 
	A different trend emerges for the \emph{Virus} dataset. Although GraphTFactor substantially improves performance for the smaller ESM models, the largest 48-layer model enables Logistic Regression to slightly surpass GraphTFactor in both Balanced Accuracy and Sensitivity. This observation suggests that viral transcription factors are largely characterized by global sequence patterns that are already effectively captured by the largest protein language model, reducing the relative benefit of graph reasoning.
	To conclude, GraphTFactor improves upon sequence-level classifiers by exploiting complementary structural information captured through graph construction and message passing. Furthermore, unlike the Logistic Regression baseline, GraphTFactor enables biologically meaningful interpretation through explainability methods such as GECo \cite{amato2026geco}, allowing the identification of relevant protein regions contributing to the prediction. This combination of performance and interpretability represents a key advantage of the proposed framework.
	
	\subsection{Excluding the graph and the GNN: ESM-2 Embeddings with Multi-Layer Perceptron}
	This experiment evaluates whether a more expressive non-linear classifier can exploit the information encoded by the ESM-2 embeddings without the need for encoding the protein as a graph. To this end, we used the mean-pooled ESM-2 representation for training and testing an MLP classifier consisting of three hidden layers with 256, 128, and 64 neurons. Each hidden layer is followed by a ReLU activation function, and the final output layer contains two output neurons (see Figure~\ref{fig:excluding_graph_GNN}). The model was optimized using the Adam algorithm \cite{KingmaB14} with a learning rate of $10^{-3}$ and trained using the cross-entropy loss function for 50 epochs with a batch size of 128. Compared to the Logistic Regression model, the MLP consistently improves performance, confirming that a non-linear decision boundary can better exploit the information encoded in the ESM-2 embeddings. On the \emph{All} dataset, BA increases from 0.904 to 0.919 with ESM-2 6 and from 0.946 to 0.961 with ESM-2 30, while Se improves from 0.840 to 0.852 and from 0.913 to 0.934, respectively. Despite this improvement, GraphTFactor remains superior, achieving BA values of 0.966 and 0.973 for the same two models. The advantage is particularly pronounced with smaller ESM models: with ESM-2 6, GraphTFactor improves BA by 4.7 percentage points and Se by 9.4 percentage points over the MLP. Even with ESM-2 48, where the MLP reaches 0.972 BA and 0.953 Se, GraphTFactor achieves 0.983 BA and 0.974 Se. A similar trend is observed for the \emph{Eukaryotic} dataset. GraphTFactor improves BA from 0.902 to 0.946 with ESM-2 6 and from 0.953 to 0.971 with ESM-2 48. The corresponding Se increases from 0.819 to 0.902 and from 0.916 to 0.955, respectively. These results indicate that the graph-based representation provides complementary information that cannot be fully recovered by applying a nonlinear classifier to a single global protein embedding. Interestingly, the Virus dataset exhibits the opposite behavior. The MLP consistently outperforms GraphTFactor, with BA increasing from 0.929 to 0.974 across ESM-2 6 to 48, compared with 0.908 to 0.951 for GraphTFactor. Se follows the same trend, reaching 0.962 for the MLP compared with 0.912 for GraphTFactor with ESM-2 48. This behavior suggests that, for the \emph{Virus} dataset, global sequence representations are already highly informative and that the additional regional structure introduced by GraphTFactor does not provide a consistent advantage. To conclude from an interpretability perspective, the MLP also presents an important limitation. Although post-hoc explainability methods can be applied to the classifier, they can only provide feature-level importance scores and do not fully recover the specific protein sequence regions responsible for the prediction. In contrast, GraphTFactor explicitly represents proteins as collections of sequence regions, enabling its predictions to be traced back to specific portions of the protein sequence. Thus, beyond the predictive performance, the graph-based formulation provides a more biologically interpretable representation, allowing the identification of potentially relevant sequence regions rather than only the most influential components of a global embedding vector.
	
	\subsection{Excluding the GNN: Graph Construction with Readout and Logistic Regression}
	To isolate the GNN's contribution within the GraphTFactor framework, we retained the full graph construction pipeline but replaced the message-passing network with a simple global graph mean readout followed by Logistic Regression.  Consequently, the generated graphs preserve the proposed MDL segmentation, region embeddings, and similarity-based edge construction, while removing all graph convolution operations (see Figure~\ref{fig:excluding_GNN}). This experiment yields one of the largest performance degradations throughout the ablation study. For the \emph{All} dataset, BA decreases from 0.966 to 0.780 when using the smallest ESM-2 model, while Se drops from 0.946 to 0.622. Although the gap progressively narrows as larger ESM-2 models are adopted, GraphTFactor consistently maintains a clear advantage for every embedding size. These findings demonstrate that graph construction alone is insufficient to achieve competitive performance. While the generated graph representations already encode meaningful biological information, the graph neural network substantially enriches them by propagating information among neighboring sequence regions and modeling higher-order dependencies that simple global pooling cannot capture. The effect is particularly pronounced for the \emph{Eukaryotic} dataset. With ESM-2 6, the readout-based approach achieves only 0.810 BA and 0.686 Se, compared with 0.946 BA and 0.902 Se for GraphTFactor. Even with ESM-2 48, the gap remains approximately 1.8 percentage points in BA and 1.9 percentage points in Sensitivity. These results show that the graph representation itself contains useful information, but simple global aggregation cannot fully exploit the relationships encoded by the graph.
	A similar trend is observed for the \emph{Prokaryotic} dataset, although the absolute differences are smaller because both approaches approach saturation with larger ESM models. For ESM-2 6, GraphTFactor improves BA from 0.865 to 0.983 and Se from 0.772 to 0.974. With ESM-2 48, the corresponding values are 0.992 vs. 0.988 for BA and 0.987 vs. 0.983 for Se. This convergence indicates that highly expressive ESM representations can partially compensate for the absence of message passing when the underlying classification problem is already relatively easy. The \emph{Virus} dataset again stands out as an exception. For ESM-2 48, the readout-based classifier achieves 0.955 BA and 0.923 Se, slightly exceeding GraphTFactor's 0.951 BA and 0.912 Se. This is consistent with the previous MLP experiment and further suggests that, for viral proteins, global sequence-level information may be sufficient for accurate classification. In contrast, the substantial improvements obtained by GraphTFactor on the \emph{All} and \emph{Eukaryotic} datasets provide direct evidence that message passing is an important component of the proposed architecture, enabling the model to exploit interactions among protein regions that a global readout alone cannot capture.
	
	\subsection{Excluding the ESM-2 embeddings: Graph Construction Using Random Amino Acid Embeddings}
	To quantify the importance of contextual protein representations learned by ESM-2, we replaced pretrained residue embeddings with randomly initialized amino acid vectors while preserving the remaining graph construction pipeline (see Figure~\ref{fig:excluding_EMS-2}). Unlike fully random residue-level features, this approach preserves the amino acid identity while removing the evolutionary and contextual information captured by the pretrained protein language model.  The graph construction pipeline was kept unchanged, including the MDL-based sequence segmentation, weighted
	aggregation of residue embeddings into region-level representations, computation of local differences, feature normalization, and similarity-based edge generation. Therefore, the resulting graphs maintain the same construction strategy while relying on non-informative residue representations that lack the semantic information learned by ESM-2. We then trained and evaluated GraphTFactor under the same experimental protocol used for the original ESM-2-based graphs. This ablation quantifies the specific contribution of pretrained protein representations beyond the information provided by amino acid composition and graph topology. Among all ablation experiments, this modification produces the largest reduction in predictive performance. Across the \emph{All} dataset, BA decreases from approximately 0.97–0.98 to approximately 0.82, while Se falls from nearly 0.97 to approximately 0.67, regardless of the embedding dimensionality. Similar reductions are consistently observed for the \emph{Eukaryotic}, \emph{Prokaryotic}, and \emph{Virus} datasets. An important observation is that increasing the dimensionality of the random embeddings from 320 to 5120 dimensions does not improve performance. This indicates that predictive accuracy is not determined by feature dimensionality itself, but rather by the biological knowledge encoded within pretrained ESM-2 representations. Although the graph topology remains identical, replacing contextual residue embeddings with random vectors removes the potential biological information necessary to distinguish transcription factors. Consequently, GraphTFactor is unable to recover the lost information through graph learning alone. These results identify pretrained protein language model representations as one of the most critical components of the proposed framework.
	
	\subsection{Excluding the MDL-based segmentation: Random Sequence Segmentation}
	Finally, we evaluated the contribution of the proposed MDL-based segmentation strategy by replacing it with a randomized segmentation procedure (see Figure~\ref{fig:excluding_MDL}). In this variant, the number of graph nodes was determined independently of the MDL optimization and adapted to the protein sequence length according to a square-root scaling. Specifically, for a protein sequence of length $n$, the number of regions was set to $K = \left\lceil \sqrt{n} \right\rceil$. The boundaries between consecutive regions were then sampled randomly along the sequence, generating alternative partitions while preserving the same length-dependent graph granularity. This approach provides a sequence-length-adaptive representation while eliminating the use of information-theoretic criteria for identifying biologically or structurally meaningful sequence regions. 
	The downstream graph construction pipeline, including ESM-based region embeddings, local feature computation, feature normalization, similarity-based edge generation, and graph neural network training, remained identical to the original framework. Consequently, the only modified component was the strategy used to partition protein sequences into graph nodes. Unlike the proposed MDL segmentation, which identifies variable-length regions by minimizing an information-theoretic description length criterion, random segmentation ignores sequence-specific statistical structure. This experiment therefore quantifies the importance of meaningful sequence partitioning for the overall performance of GraphTFactor. Compared with the complete GraphTFactor framework, random segmentation consistently reduces both BA and Se across all datasets and embedding sizes. The degradation is particularly evident for smaller ESM-2 models, where BA decreases by approximately three percentage points on the \emph{All} dataset. As larger language models are employed, the performance gap becomes smaller, suggesting that richer residue representations partially compensate for suboptimal region boundaries. Nevertheless, the proposed MDL segmentation consistently remains superior throughout all experimental settings. Interestingly, the reduction introduced by random segmentation is noticeably smaller than that observed when replacing pretrained embeddings with random vectors or removing message passing entirely. This finding suggests that, although meaningful sequence partitioning contributes positively to graph quality, the biological information encoded by ESM-2 embeddings and the GNN itself have an even greater influence on final predictive performance. Nevertheless, the consistent superiority of MDL segmentation demonstrates that identifying statistically coherent protein regions produces more informative graph nodes than arbitrary sequence partitions, ultimately facilitating downstream graph learning.
	
	\subsection{Discussion}
	Overall, the ablation study demonstrates that GraphTFactor derives its predictive performance from the interaction of several complementary components rather than from a single design choice. Pretrained ESM-2 embeddings provide rich contextual representations that constitute the primary source of biological information, while the proposed graph construction transforms these residue-level representations into meaningful region-level graphs. The GNN subsequently exploits the relationships among similar regions through message passing, producing substantial improvements over classifiers operating on either global protein embeddings or static graph representations. Finally, the proposed MDL-based segmentation consistently generates higher-quality graph nodes than randomized sequence partitioning, further improving downstream classification.
	Taken together, these experiments demonstrate that none of the individual components alone is sufficient to reproduce the performance of the complete GraphTFactor architecture. Instead, combining contextual protein language model representations, information-theoretic sequence segmentation, biologically motivated graph construction, and graph neural network learning enables GraphTFactor to consistently achieve the highest performance across most datasets considered in this study. 
	
	\begin{landscape}
		\begin{table}[t]
			\centering
			\caption{\emph{All} dataset: performance of GraphTFactor and the considered baseline and ablation configurations across different ESM-2 layers. Results are reported in terms of Balanced Accuracy and Sensitivity. Cell colors indicate the ranking of the results: \colorbox{best}{best}, \colorbox{second}{second-best}, and \colorbox{third}{third-best}.}
			\label{tab:ablation_all}

			\resizebox{\linewidth}{!}{
				\begin{tabular}{lcccccc}
					\toprule
					
					\multicolumn{7}{c}{\textbf{Balanced Accuracy}} \\
					\midrule
					\textbf{Exp / ESM-2 Layers}
					& \textbf{6}
					& \textbf{12}
					& \textbf{30}
					& \textbf{33}
					& \textbf{36}
					& \textbf{48} \\
					
					\midrule
					
					GraphTF
					& \cellcolor{best}$0.966 \pm 0.002$ & \cellcolor{best}$0.970 \pm 0.002$ & \cellcolor{best}$0.973 \pm 0.002$ & \cellcolor{best}$0.979 \pm 0.001$ & \cellcolor{best}$0.982 \pm 0.001$ & \cellcolor{best}$0.983 \pm 0.001$ \\
					
					ESM-2 + LR
					& $0.904 \pm 0.004$ & $0.934 \pm 0.015$ & $0.946 \pm 0.003$ & $0.962 \pm 0.002$ & \cellcolor{second}$0.974 \pm 0.002$ & \cellcolor{second}$0.980 \pm 0.001$ \\
					
					ESM-2 + MLP
					& \cellcolor{third}$0.919 \pm 0.007$ & \cellcolor{third}$0.941 \pm 0.004$ & \cellcolor{second}$0.961 \pm 0.004$ & \cellcolor{second}$0.969 \pm 0.004$ & \cellcolor{third}$0.972 \pm 0.003$ & $0.972 \pm 0.004$ \\

					Graph Readout + LR
					& $0.780 \pm 0.006$ & $0.843 \pm 0.006$ & $0.868 \pm 0.006$ & $0.922 \pm 0.002$ & $0.949 \pm 0.004$ & $0.964 \pm 0.002$ \\
					
					GraphTF with random AA emb.
					& $0.829 \pm 0.017 $ & $0.826 \pm 0.016$ & $0.826 \pm 0.018$ & $0.825 \pm 0.014$ & $0.823 \pm 0.016$ & $0.824 \pm 0.011$ \\
					
					GraphTF with random seq. seg.
					& \cellcolor{second}$0.937 \pm 0.003$  & \cellcolor{second}$0.949 \pm 0.003$ & \cellcolor{third}$0.955 \pm 0.003$ & \cellcolor{third}$0.966 \pm 0.002$ & $0.970 \pm 0.001$ & \cellcolor{third}$0.972 \pm 0.003$ \\
					
					\midrule
					\multicolumn{7}{c}{\textbf{Sensitivity}} \\
					\midrule
					
					\textbf{Exp / ESM-2 Layers}
					& \textbf{6}
					& \textbf{12}
					& \textbf{30}
					& \textbf{33}
					& \textbf{36}
					& \textbf{48} \\
					
					\midrule
					
					GraphTF
					& \cellcolor{best}$0.946 \pm 0.005$ & \cellcolor{best}$0.953 \pm 0.003$ & \cellcolor{best}$0.955 \pm 0.004$ & \cellcolor{best}$0.966 \pm 0.004$ & \cellcolor{best}$0.973 \pm 0.003$ & \cellcolor{best}$0.974 \pm 0.004$ \\
					
					ESM-2 + LR
					& $0.840 \pm 0.001$ & $0.892 \pm 0.002$ & $0.913 \pm 0.006$ &\cellcolor{third} $0.939 \pm 0.005$ & \cellcolor{second}$0.959 \pm 0.004$ & \cellcolor{second}$0.970 \pm 0.002$ \\
					
					ESM-2 + MLP
					& \cellcolor{third}$0.852 \pm 0.015$ &\cellcolor{third} $0.894 \pm 0.009$ & \cellcolor{second}$0.934 \pm 0.008$ & $0.950 \pm 0.009$ & \cellcolor{third}$0.958 \pm  0.004$ & $0.953 \pm 0.010$ \\
					
					Graph Readout + LR
					& $0.622 \pm 0.012$ & $0.737 \pm 0.009$ & $0.782 \pm 0.011$ & $0.876 \pm 0.005$ & $0.925 \pm 0.006$ & $0.951 \pm 0.004$ \\
					
					GraphTF with random AA emb.
					& $0.682 \pm 0.044$ & $0.674 \pm 0.038$ & $0.675 \pm 0.044$ & $0.673 \pm 0.034$ & $0.667 \pm 0.023$ & $0.670 \pm  0.027$ \\
					
					GraphTF with random seq. seg.
					& \cellcolor{second}$0.898 \pm 0.006$ & \cellcolor{second}$0.916 \pm 0.006$ & \cellcolor{third}$0.926 \pm 0.009$ & \cellcolor{second}$0.945 \pm 0.007$ & $0.950 \pm 0.004$ &\cellcolor{third} $0.953 \pm 0.006$ \\
					
					\bottomrule
				\end{tabular}
			}
		\end{table}
	\end{landscape}

	\begin{landscape}
		\begin{table}[t]
			\centering
			\caption{\emph{Eukaryotic} dataset : performance of GraphTFactor and the considered baseline and ablation configurations across different ESM-2 layers. Results are reported in terms of Balanced  Accuracy and Sensitivity. Cell colors indicate the ranking of the results: \colorbox{best}{best}, \colorbox{second}{second-best}, and \colorbox{third}{third-best}.}
			\label{tab:ablation_euk}
			
			\resizebox{\linewidth}{!}{
				\begin{tabular}{lcccccc}
					\toprule
					
					\textbf{Experiment / ESM-2 Layer}
					& \textbf{6}
					& \textbf{12}
					& \textbf{30}
					& \textbf{33}
					& \textbf{36}
					& \textbf{48} \\
					
					\midrule
					\multicolumn{7}{c}{\textbf{Balanced Accuracy}} \\
					\midrule
					
					GraphTF
					& \cellcolor{best}$0.946 \pm 0.006$ & \cellcolor{best}$0.955 \pm0.005$ & \cellcolor{best}$0.959 \pm 0.005$ & \cellcolor{best}$0.963 \pm 0.004$ & \cellcolor{best}$0.968 \pm 0.005$ & \cellcolor{best}$0.971 \pm 0.003$ \\
					ESM-2 + LR
					& $0.903 \pm 0.005$ & $0.925 \pm 0.006$ & \cellcolor{third}$0.934 \pm 0.006$ & \cellcolor{third}$0.943 \pm 0.004$  & \cellcolor{second}$ 0.956 \pm 0.003$ & \cellcolor{second}$0.968 \pm 0.002$\\
					ESM-2 + MLP
					& \cellcolor{third}$0.907 \pm 0.013$ & \cellcolor{second}$0.930 \pm 0.010$ & \cellcolor{second} $0.942 \pm 0.011$ & \cellcolor{second}$0.946 \pm 0.008$ & \cellcolor{third}$0.953 \pm 0.012$ & $0.952 \pm 0.016$ \\
					Graph Readout + LR
					& $0.810 \pm 0.006$& $0.850 \pm 0.007$ & $0.876 \pm 0.006$ & $0.911 \pm o.006$ & $0.929 \pm 0.006$ &\cellcolor{third}$0.953 \pm 0.003$ \\
					GraphTF with random AA emb.
					&$0.797 \pm 0.012$ & $0.803 \pm 0.019$ & $0.806 \pm 0.009$ & $0.796 \pm 0.014$ & $0.803 \pm 0.021$ & $0.805 \pm 0.026$ \\
					GraphTF with random seq. seg.
					& \cellcolor{second}$0.909 \pm 0.008$ & \cellcolor{third}$0.927 \pm 0.018$ & $0.932 \pm 0.008$ & $0.943 \pm 0.008$ & $0.947 \pm 0.007$ & $0.951 \pm 0.008$ \\
					
					\midrule
					\multicolumn{7}{c}{\textbf{Sensitivity}} \\
					\midrule
					
					GraphTF
					& \cellcolor{best}$0.902 \pm 0.017$ & \cellcolor{best}$0.931 \pm 0.009$ & \cellcolor{best}$0.931 \pm 0.012$ & \cellcolor{best}$0.938 \pm 0.008$ & \cellcolor{best}$0.949 \pm 0.010$ & \cellcolor{best}$0.955 \pm 0.007$ \\
					ESM-2 + LR
					& \cellcolor{third}$0.834 \pm 0.011$ & \cellcolor{third}$0.874 \pm 0.013$ & \cellcolor{third}$0.889 \pm 0.013$ & \cellcolor{third}$0.904 \pm 0.009$ & \cellcolor{second}$0.928 \pm 0.007$  & \cellcolor{second}$0.951 \pm 0.005$\\
					ESM-2 + MLP
					& $0.819 \pm 0.027$ & $0.874 \pm 0.022$ & \cellcolor{second}$0.898 \pm 0.025$ & \cellcolor{second}$0.907 \pm 0.020$ & \cellcolor{third}$0.920 \pm 0.027$ & $0.916 \pm 0.034$ \\
					Graph Readout + LR
					& $0.686 \pm 0.002$  & $0.754 \pm 0.013$ & $0.798 \pm 0.015$ & $0.861 \pm 0.013$ & $0.900 \pm 0.009$ & \cellcolor{third}$0.936 \pm 0.004$ \\
					GraphTF with random AA emb.
					& $0.614 \pm 0.027$ & $0.624 \pm 0.042$ & $0.632 \pm 0.021$ & $0.610 \pm 0.032$ & $0.626 \pm 0.046$ & $0.631 \pm 0.058$ \\
					GraphTF with random seq. seg.
					& \cellcolor{second}$0.843 \pm 0.021$ & \cellcolor{second}$0.878 \pm 0.021$ & $0.884 \pm 0.020$ & $0.904 \pm 0.022$ & $0.910 \pm 0.018$ & $0.917 \pm 0.019$\\
					
					\bottomrule
				\end{tabular}
			}
		\end{table}
	\end{landscape}

	\begin{landscape}
		\begin{table}[t]
			\centering
			\caption{\emph{Prokaryotic} dataset : performance of GraphTFactor and the considered baseline and ablation configurations across different ESM-2 layers. Results are reported in terms of Balanced  Accuracy and Sensitivity. Cell colors indicate the ranking of the results: \colorbox{best}{best}, \colorbox{second}{second-best}, and \colorbox{third}{third-best}.}
			\label{tab:ablation_prok}
			
			\resizebox{\linewidth}{!}{
				\begin{tabular}{lcccccc}
					\toprule
					
					\textbf{Experiment / ESM-2 Layer}
					& \textbf{6}
					& \textbf{12}
					& \textbf{30}
					& \textbf{33}
					& \textbf{36}
					& \textbf{48} \\
					
					\midrule
					\multicolumn{7}{c}{\textbf{Balanced Accuracy}} \\
					\midrule
					
					GraphTF
					& \cellcolor{best}$0.988 \pm 0.001$ & \cellcolor{best}$0.991 \pm0.001$ & \cellcolor{best}$0.993 \pm 0.001$ & \cellcolor{ best}$0.994 \pm 0.004$ & \cellcolor{best}$0.998 \pm 0.001$ \cellcolor{best}& \cellcolor{best}$0.994 \pm 0.001$ \\
					ESM-2 + LR
					& $0.942 \pm 0.001$& \cellcolor{third}$0.968 \pm 0.003$ & \cellcolor{third} $0.976 \pm 0.002$ & \cellcolor{second}$0.987 \pm 0.003$ &\cellcolor{second} $0.991 \pm  0.002$  & \cellcolor{second}$0.993 \pm 0.001$ \\
					ESM-2 + MLP
					& \cellcolor{third}$0.947 \pm 0.010$ & $0.955 \pm 0.008$ & $0.954 \pm 0.013$ &$0.956 \pm 0.014$ & $0.987 \pm 0.007$ & \cellcolor{third}$0.990 \pm 0.003$\\
					Graph Readout + LR
					& $0.865 \pm 0.008$ & $0.927 \pm 0.006$ & $0.936 \pm 0.003$ & $0.968 \pm 0.004$ & $0.984 \pm 0.001$ & $0.988 \pm 0.002$ \\
					GraphTF with random AA emb.
					& $0.890 \pm 0.007$ & $0.893 \pm 0.007$ & $0.894 \pm 0.006$ & $0.892 \pm 0.006$ & $0.888 \pm 0.008$ & $0.892 \pm 0.008$ \\
					GraphTF with random seq. seg.
					& \cellcolor{second}$0.967 \pm 0.005$   & \cellcolor{second}$0.972 \pm 0.003$ & \cellcolor{second} $0.979 \pm 0.002$ & $\cellcolor{third}0.985 \pm 0.002$ & \cellcolor{third}$0.989 \pm 0.002$ & $0.988 \pm 0.003$ \\
					
					\midrule
					\multicolumn{7}{c}{\textbf{Sensitivity}} \\
					\midrule
					
					GraphTF
					& \cellcolor{best}$0.974 \pm 0.006$ & $\cellcolor{best}0.980 \pm 0.004$ & \cellcolor{best}$0.984 \pm 0.005$ & \cellcolor{best} $0.986 \pm 0.003$ & \cellcolor{best}$0.989 \pm 0.003$ & \cellcolor{second}$0.987 \pm 0.002$ \\
					ESM-2 + LR
					& \cellcolor{third}$0.907 \pm 0.008$ & \cellcolor{third}$0.949 \pm 0.007$ & \cellcolor{second}$0.965 \pm 0.005$ & \cellcolor{second}$0.980 \pm 0.005$ & \cellcolor{second}$0.987 \pm 0.004$ & \cellcolor{best}$0.989 \pm 0.003$ \\
					ESM-2 + MLP
					& $0.903 \pm 0.020$ & $0.916 \pm 0.017$ & \cellcolor{third}$0.912 \pm 0.027$ & $0.916 \pm 0.081$ & $0.979 \pm 0.014$ & $0.982 \pm 0.005$ \\
					Graph Readout + LR
					& $0.771 \pm 0.015$ & $0.880 \pm 0.013$ & $0.896 \pm 0.005$ & $0.955 \pm 0.007$ & $0.978 \pm 0.003$ & $0.983 \pm 0.003$ \\
					GraphTF with random AA emb.
					& $0.821 \pm 0.028$ & $0.828 \pm 0.024$ & $0.840 \pm 0.019 $ & $0.830 \pm 0.027$ & $0.820 \pm 0.032$ & $0.831 \pm 0.030$ \\
					GraphTF with random seq. seg.
					& \cellcolor{second}$0.946 \pm 0.014$ & \cellcolor{second}$0.952 \pm 0.016$ & \cellcolor{second} $0.965 \pm 0.005$ & \cellcolor{third} $0.977 \pm 0.003$ & \cellcolor{third}$0.982 \pm 0.004$ & \cellcolor{third}$0.981 \pm 0.005$  \\
					
					\bottomrule
				\end{tabular}
			}
		\end{table}
	\end{landscape}
	
	\begin{landscape}
		\begin{table}[t]
			\centering
			\caption{\emph{Virus} dataset: performance of GraphTFactor and the considered baseline and ablation configurations across different ESM-2 layers. Results are reported in terms of Balanced  Accuracy and Sensitivity. Cell colors indicate the ranking of the results: \colorbox{best}{best}, \colorbox{second}{second-best}, and \colorbox{third}{third-best}.}
			\label{tab:ablation_vir}
			
			\resizebox{\linewidth}{!}{
				\begin{tabular}{lcccccc}
					\toprule
					
					\textbf{Experiment / ESM-2 Layer}
					& \textbf{6}
					& \textbf{12}
					& \textbf{30}
					& \textbf{33}
					& \textbf{36}
					& \textbf{48} \\
					
					\midrule
					\multicolumn{7}{c}{\textbf{Balanced Accuracy}} \\
					\midrule
					
					GraphTF
					& \cellcolor{best}$0.931 \pm 0.012$ & \cellcolor{best}$0.953 \pm0.010$ & \cellcolor{best}$0.954 \pm 0.010$ & \cellcolor{best} $0.967 \pm 0.009$ & \cellcolor{second}$0.966 \pm 0.018$ & \cellcolor{third}$0.971 \pm 0.007$ \\
					ESM-2 + LR
					& $0.818 \pm 0.033$ & $0.826 \pm 0.032$ & $0.841 \pm 0.034$ & $0.882 \pm 0.030$ & $0.926 \pm 0.026$ & \cellcolor{best} $0.974  \pm 0.017$\\
					ESM-2 + MLP
					& \cellcolor{second}$0.929 \pm 0.015$ & \cellcolor{second}$0.947 \pm 0.016$ & \cellcolor{second}$0.951 \pm 0.014$ & \cellcolor{second}$0.951 \pm 0.014$ & \cellcolor{best} $0.972 \pm 0.011$ &  \cellcolor{second}$0.973 \pm 0.013$ \\
					Graph Readout + LR
					& $0.830 \pm 0.036$ & \cellcolor{third}$0.875 \pm 0.020$ & $0.879 \pm 0.019$ & \cellcolor{third} $0.931 \pm 0.018$ & \cellcolor{third} $0.947 \pm 0.013$ & $0.955 \pm 0.014$ \\
					GraphTF with random AA emb.
					& $0.827 \pm 0.006$ & $0.842 \pm 0.027$ & $0.828 \pm 0.035 $ & $0.839 \pm 0.027$ & $0.824 \pm 0.029$ & $0.820 \pm 0.029$ \\
					GraphTF with random seq. seg.
					&   \cellcolor{third}$0.831 \pm 0.021$ & $0.850 \pm 0.026$ & \cellcolor{third}$0.896 \pm 0.025$ & $0.883 \pm 0.032$ & $0.905 \pm 0.029$ & $0.910 \pm 0.028$ \\
					
					\midrule
					\multicolumn{7}{c}{\textbf{Sensitivity}} \\
					\midrule
					
					GraphTF
					& \cellcolor{second}$0.862 \pm 0.059$ & \cellcolor{second}$0.881 \pm 0.053$ & \cellcolor{second}$0.880 \pm 0.049$ & \cellcolor{second}$0.908 \pm 0.034$ & \cellcolor{second}$0.914 \pm 0.037$ & \cellcolor{second}$0.912 \pm 0.021$ \\
					ESM-2 + LR
					& $0.684 \pm 0.067$ & $0.695 \pm 0.060$ & $0.719 \pm 0.067$ & $0.788 \pm 0.058$ & $0.858 \pm 0.051$ & $0.955 \pm 0.030$ \\
					ESM-2 + MLP
					& \cellcolor{best}$0.899 \pm 0.045$ & \cellcolor{best}$0.929 \pm 0.030$ & \cellcolor{best}$0.936 \pm 0.021$ & \cellcolor{best}$0.930 \pm 0.031$ & \cellcolor{best}$0.962 \pm 0.021$ & \cellcolor{best}$0.962 \pm 0.024$ \\
					Graph Readout + LR
					& \cellcolor{third}$0.739 \pm 0.062$ & \cellcolor{third}$0.830 \pm 0.030$ & \cellcolor{third}$0.825 \pm 0.041$ & \cellcolor{third}$0.892 \pm 0.031$ & \cellcolor{third}$0.912 \pm 0.024$ & \cellcolor{third}$0.923 \pm 0.028$ \\
					GraphTF with random AA emb.
					& $0.732 \pm 0.079$ & $0.758 \pm 0.080$ & $0.739 \pm 0.082$ & $0.749 \pm 0.060$ & $0.745 \pm 0.080$ & $0.732 \pm 0.075$ \\
					GraphTF with random seq. seg.
					& $0.721 \pm 0.052$ & $0.756 \pm 0.056$ & $0.777 \pm 0.064$ & $0.797 \pm 0.059$ &  $0.821 \pm 0.063$ & $0.832 \pm 0.058$ \\
					
					\bottomrule
				\end{tabular}
			}
		\end{table}
	\end{landscape}
\end{appendix}

\bibliographystyle{plain}
\bibliography{biblio.bib}

\end{document}